\documentclass{WileyMSP-template}

\usepackage{ragged2e}
\AtBeginDocument{
  \justifying
  \setlength{\parindent}{0pt}
}
\usepackage{xcolor}
\usepackage{amsmath}

\begin{document}

\pagestyle{fancy}
\rhead{\includegraphics[width=2.5cm]{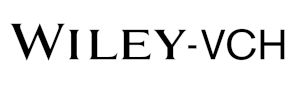}}

\title{Voltage-Controlled Dielectric Cavity Magnonics in Epitaxial Fe Films on SrTiO$_3$ Single Crystal Substrates}

\maketitle

\author{Stijn Reniers$^{1}$*}
\author{Ekaterina Gorokh$^{1}$}
\author{Lukas Nulens$^{1}$}
\author{Andries Boelen$^{2,3}$}
\author{Christian Haffner$^{3}$}
\author{Clement Merckling$^{3,2}$}
\author{Alejandro V. Silhanek$^{4}$}
\author{Kristiaan Temst$^{1,3}$}
\author{Joris Van de Vondel$^{1}$*}

\begin{affiliations}

$^{1}$ Quantum Solid-State Physics, Department of Physics and Astronomy, KU Leuven, Celestijnenlaan 200D, Leuven B-3001, Belgium

$^{2}$Department of Materials Engineering (MTM), KU Leuven, B-3001 Leuven, Belgium

$^{3}$Imec, Kapeldreef 75, Leuven, Belgium

$^{4}$Experimental Physics of Nanostructured Materials, Q-MAT, Department of Physics, Université de Liège, Liège B-4000, Belgium

*stijn.reniers@kuleuven.be
*joris.vandevondel@kuleuven.be
\end{affiliations}


\keywords{Cavity Magnonics, Voltage Controlled Magnetism, Microwave technology}

\begin{abstract}
Cavity magnonics relies on the hybridization between collective spin excitations and well-defined photonic or mechanical cavity modes, enabling fast and efficient coupling of spintronic devices to microwave circuits and optical information carriers. In this work, we present a hybrid material heterostructure that combines an intrinsic dielectric microwave cavity with a ferromagnetic medium, establishing magnon-photon interaction with coupling rates in the order of 0.1-1 GHz. Demonstrating effective voltage-control of the hybrid mode spectrum at cryogenic temperatures, we achieve a highly tunable photo-magnonic material, that could provide significant advancements for high-performance and power-efficient cryogenic RF-applications, including hybrid quantum technology, superconducting electronics and nonlinear microwave devices for unconventional computation and signal processing. \end{abstract}

\section{Introduction}\label{Intro}

\textit{Cavity magnonics} forms an evolving field of research, that studies the controlled interaction between collective spin excitations (magnons) and confined electromagnetic waves (cavity photons) \cite{ZARERAMESHTI20221,Harder21, Macedo_2025}. Such controlled magnon-photon coupling enables microwave logic functionality for spintronics, telecommunication, sensing technology and solid-state quantum computing \cite{ZHANGReview, Lachance-Quirion_2019, GolovchanskiySFHybrid, AFMGolovchanskiy, Kurebayashi2026, TABUCHI2016729, QuantumMagnonics26} (cavity electromagnonics), as well as upconversion from microwave- to optical frequencies \cite{Rueda:16, Fan2018Superconducting, Holzgrafe:20, McKenna:20, Optomagnonics_mag_sol, ZhuCavityOptomagnonics20} (cavity optomagnonics), providing a crucial functional interface between cryogenic quantum processors and optical communication networks.

\vspace{5 pt}

Traditionally, cavity magnonics relies on direct magnetic dipole coupling between magnons and photons confined in \textit{closed}, high-Q cavities \cite{Huebl13, PirroCoherent}. This configuration enables coherent and low-loss energy exchange, making it particularly attractive for hybrid quantum systems and applications where the preservation of coherence is essential \cite{TABUCHI2016729, TabuchiYIGsphere, PRX_QuantumMagnonics, Candido_quantum_2021}. Alternatively, \textit{open} cavity configurations couple magnons to microwave photons through engineered electromagnetic environments, such as transmission lines and planar resonators \cite{Harder21, WangMing_Dissipative_cavitymagnonics}. Such architectures naturally introduce radiative and dissipative channels, but substantially relax the geometric and integration constraints associated with maintaining closed, high-Q cavities \cite{BhoiandBosung19, Harder18, Yang24, MetelmannNonReciprocal, RaoDissipative}. As such, they offer greater flexibility for planar integration, device miniaturization, and direct interfacing with microwave circuitry, making them particularly attractive for scalable implementations with enhanced external tunability. Practical, circuit-level implementations would likely rely on a combination of open and closed cavity systems, trading off cavity coherence with miniaturization and circuit process margins.

\vspace{5 pt}

Substantial progress in both open- and closed cavity magnonics has been realized in macroscopic prototype systems. Early works rely on relatively large 3-dimensional microwave cavities, with a size of several cm \cite{TabuchiYIGsphere, LiHybridMagnonics, ZhangYIGsphere14, BourcinG}, using the ferrimagnetic insulator Yttrium Iron Garnet (YIG) as magnonic medium, which exhibits low magnetic damping and  optical losses. While smaller-scale implementations based on YIG thin films have shown important novel functionality for both electro- \cite{YIGSpinWaves, SCMagnonicsYIGYBCO, Santos2023, Grishin26YIGNLdevice} and optomagnonics \cite{ZhuCavityOptomagnonics20}, the challenging growth procedure on Gadolinium Gallium Garnet (GGG) substrates and stronger damping at low temperatures limit the intrinsic scalability and compatibility with nanoscale cryogenic circuits \cite{YIG_thin_film_challenges, YIG_thin_film_anisotropy, SERHA2025_YIG}. More recent works have established a `second generation' of hybrid cavity magnonics, based on planar superconducting resonators combined with low Gilbert damping ($\alpha \approx$ 1 $\cdot$ 10$^{-3}$) metallic ferromagnetic materials such as Permalloy \cite{GolovchanskiySFHybrid, Yi_SF_hybrids, MagnonicsMetamaterials} or FeCo \cite{NIST_sF_hybrids, Edwards19}. Despite generally larger magnon damping compared to YIG ($\alpha_{YIG} \approx$ 1 $\cdot$ 10$^{-4}$), the benefits of using ferromagnetic metals include larger saturation magnetization ($M_s$) and hence stronger coupling ($g \propto \sqrt{M_s}$) to microwave- and optical photons up to GHz rates, as well as enhanced compatibility with low temperatures and standard nanofabrication processes \cite{QuantumMagnonics26}. However, these innovative implementations are still geometrically constrained by the resonator design, yielding typical device footprints in the order of  millimeters. Despite these advancements, achieving nanoscale integration with enhanced miniaturization and tunability remains a key challenge.

\vspace{5 pt}

A powerful approach to pursue the miniaturization evolution towards nanoscale integration consists in realizing an intrinsic microwave cavity in a dielectric substrate layer, combined with a thin film metallic ferromagnet, forming a functional material heterostructure \cite{prePrintMagnonPhonon}. The resulting, compact hybrid \textit{magnonic cavity material} combines low microwave- and optical losses in the dielectric with the scalability and cryogenic compatibility of ferromagnetic metals. Such \textit{in-materia} cavity functionality can support significant reductions in design complexity, device footprint and alignment constraints when patterning the ferromagnetic thin film into nanoscale magnonic devices for circuit integration. The magnon–photon interaction can then be defined at the material level, while engineering of the magnonic- and electromagnetic circuitry controls the balance between coherent coupling, dissipation, and signal accessibility \cite{Harder21, Shuai2025, mukhopadhyay2026}. In such systems, the magnon-photon interaction of each circuit segment to the material cavity can be tuned by local tuning of the material properties through external control variables. 

\vspace{10 pt}

A particularly attractive material to provide such intrinsic cavity functionality is Strontium Titanate (SrTiO$_3$, referred to as `STO' in the remainder of the text), one of the most widely studied correlated oxide materials in solid-state physics. Its dielectric permittivity value $\epsilon_r$ is strongly tunable by temperature, defect chemistry, strain and electric field \cite{Muller79, Haeni2004, Sidoruk2010, Rowley2014}. The latter in particular is highly convenient, allowing voltage-tunable control of the dielectric cavity properties. Both the permittivity value itself and the impact of external electric field (i.e., gate-tunability) become very large at deep cryogenic temperatures \cite{DelftResonatorsOnSTO}, making STO an appealing functional oxide for gate-tunable electro-optic conversion and non-linear amplification for quantum computing and superconducting electronics \cite{HaffnerSciencePaper,VanGasse_electro_optic, Kamal_STO_permittivity}. Additionally, STO forms an excellent growth template for the epitaxial deposition of metallic ferromagnets, enabling high crystallinity and sharp interfaces, in particular for Fe \cite{Kamaratos_2008, Chien12} and Co \cite{epiCoSTO}. Such high-quality films exhibit low magnetic damping and are favorable to harness magnetic properties induced by structural symmetries, like magnetocrystalline- and interface anisotropies.

\vspace{10 pt}

\begin{figure*}[h]
    \centering
    \includegraphics[width=0.9\linewidth]{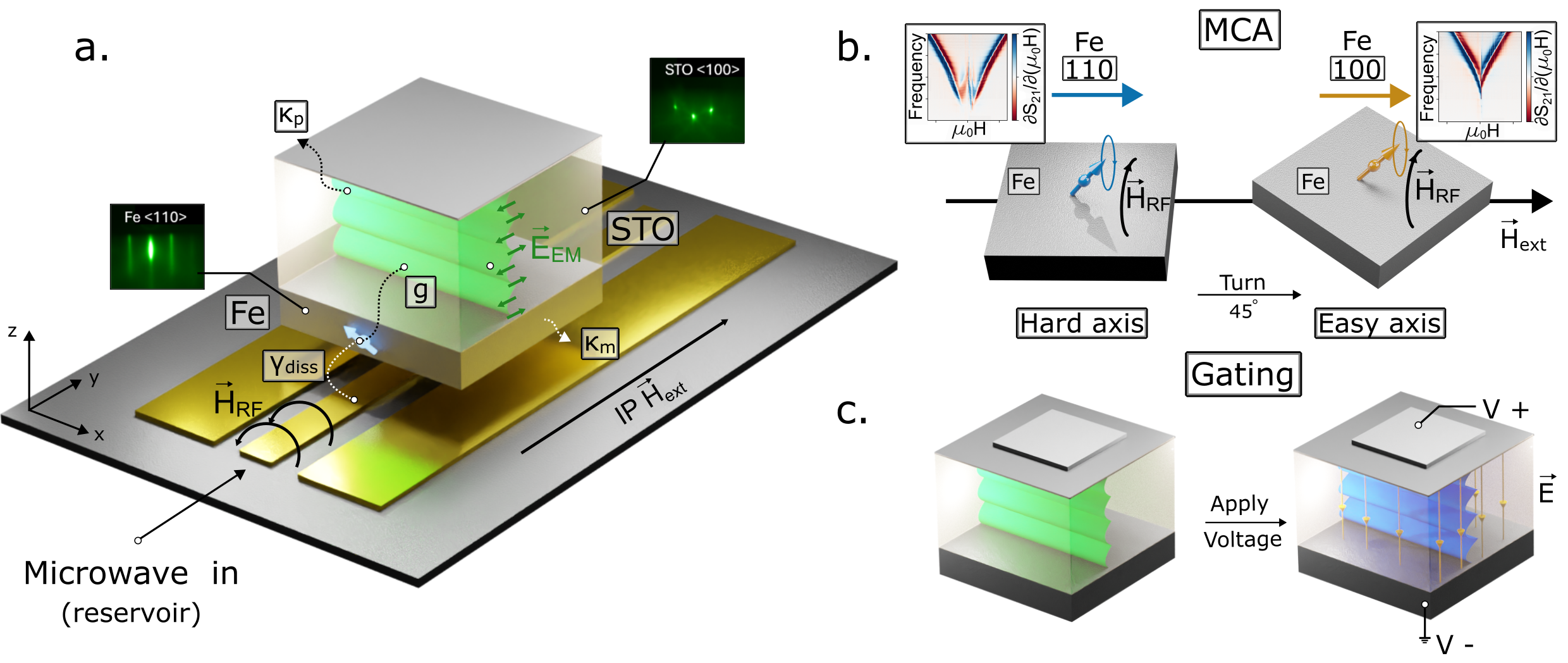}
    \caption{(a) Schematic representation of the STO/Fe magnonic cavity heterostructure, in the Fe-side excitation configuration.
    $\kappa_p$ and $\kappa_m$ represent the photon- and magnon damping rates, respectively, while $\gamma_{diss}$ and $g$ are the respective dissipative and coherent coupling rates. The light-green plane wave represents a standing microwave mode in the STO crystal. For all measurements, the in-plane (IP) static magnetic field was applied along the long axis of the coplanar waveguide. (b) Magnetocrystalline anisotropy (MCA) defines a directional magnon dispersion, and a substantial ferromagnetic resonance frequency at $\mu_0H$ = 0 mT. The insets show ferromagnetic resonance spectra recorded along the respective in-plane Fe crystalline directions. (c) Gate voltage application modifies the STO dielectric permittivity, and hence the cavity mode spectrum, shown schematically as a change of the mode frequency.}
    \label{fig:mainconcept}
\end{figure*}

In this work, we implement a magnon-photon cavity material by the epitaxial overgrowth of a ferromagnetic Fe film on a single-crystal STO (001) substrate, shown conceptually in Figure \ref{fig:mainconcept}a. The material stack is placed on a coplanar waveguide for microwave characterization. The epitaxial quality of the Fe film establishes a cubic magnetocrystalline anisotropy (MCA) energy $K_c$ exceeding 10 kJ/m$^3$ (Figure \ref{fig:mainconcept}b), substantially larger than what is attainable in epitaxial YIG films \cite{YoshimotoYIG} or polycrystalline metallic ferromagnets ($K_c<$ 1 kJ/m$^3$). The fourfold symmetric magnetic energy landscape of cubic MCA materials is reported to provide beneficial properties for spin-wave propagation \cite{Sekiguchi2017}, optomagnonic coupling \cite{Xie2025} and magnonic crystal engineering \cite{MCAMagnonicCrystals}. Furthermore, the MCA provides an effective internal \textit{anisotropy field} that induces ferromagnetic resonance in the GHz range, even in the absence of external magnetic bias field. Due to the significant circuit footprint associated with on-chip magnetic field generation, field-free operation is an important advantage considering scalability and the reduction of crosstalk in magnonic device control. It is also beneficial for potential interfacing with superconducting qubits, since external magnetic fields can strongly reduce qubit coherence times \cite{QuantumMagnonics26} and induce substantial losses in the readout circuits \cite{Nulens2023}. Within the STO/Fe hybrid material system, we exploit the STO dielectric bulk resonances to establish well-defined photonic modes, which couple effectively to the ferromagnetic resonance (FMR) modes of the Fe film when excited by an adjacent microwave transmission stripline. Moreover, the STO resonances, and hence the hybrid modes, are tunable by gate voltage (Figure \ref{fig:mainconcept}c), since the STO permittivity has a pronounced electric field dependence. This voltage control of the cavity response provides additional functionality towards a tunability scheme different from magnetic field or local current application. As such, the STO/Fe heterostructure forms a magnonic cavity material in which the spin and photonic degree of freedom are coupled, and where this coupling is strongly tunable by temperature, crystal orientation and gate voltage. 

\section{\label{sec:level2}Sample fabrication and characterization}

Single crystalline STO (001) substrates ($l = w =$ 10 mm, $t=$ 0.5 mm) were heated in Ultra-High Vacuum (UHV, $P = 1 \times 10^{-9}$ mbar) environment at 300$^\circ$C for 20 minutes. A 30 nm Fe film was then deposited epitaxially on the STO substrates by Molecular Beam Epitaxy (MBE, e-gun deposition) at a growth rate of 0.2 \AA/s and a substrate temperature of 200$^\circ$C under UHV conditions ($P = 7 \times 10^{-10}$ mbar). The sample was then post-annealed at 360$^\circ$C to improve the crystallinity of the film. RHEED patterns recorded before and after Fe deposition and the subsequent annealing (Supplementary Figure S1b) reveal the crystalline surface structure of the STO substrate and Fe layer. A 6 nm Ti capping was deposited at room temperature to protect the Fe film from oxidation in ambient environment. The exact same procedure was followed for the deposition of Fe on a Si(001) substrate/STO(10 nm) heterostructure for the reference measurement, discussed in Section \ref{characterization}. For the additional Fe/Nb sample, discussed in Section \ref{MPcoupling:coherent}, a 85 nm Nb layer was deposited at 0.8 \AA/s on top of the Fe layer (without additional capping). Further structural characterization of the Fe film by X-Ray Diffraction, shown in Supplementary Figure S1c, reveals the well-defined epitaxial relationship STO$\langle100\rangle$ $||$ Fe$\langle110\rangle$, leading to a minimized expected lattice mismatch of 3\% \cite{Kamaratos_2008}. These observations confirm the high structural quality of the STO/Fe heterostructure. The magnetic properties of the Fe film were characterized by Vibrating Sample Magnetometry (VSM) along two different azimuthal directions, shown in Supplementary Figure S2. From the measured magnetic moment and the estimated film volume, the saturation magnetization at low temperatures was determined to be (17 $\pm$ 2) $\cdot 10^2$  kA/m, close to the expected value of bulk Fe (1780 kA/m) \cite{CullityGraham}. The cubic magnetocrystalline anisotropy energy constant $K_c$ was estimated as 32 $\pm$ 4 kJ/m$^3$ from the area between the easy- and hard-axis hysteresis loops, lower but comparable to the Fe bulk value (48 kJ/m$^3$)\cite{CullityGraham}. This corresponds to a magnetic anisotropy field estimate of $H_k = \frac{2K_c}{\mu_0M_s} \approx 37 \space$ mT. Details are provided in Supplementary Section 2.

\section{Characterization of the STO cavity and Fe magnon subsystems} \label{characterization}

Microwave characterization of the STO/Fe heterostructures was performed in a Dynacool PPMS cryostat, using the cryo-FMR option. The magnonic cavity system under study consists of three interacting components: (1) the coplanar waveguide stripline on the microwave PCB, serving as the RF excitation source and electromagnetic coupling environment, (2) the STO substrate, which acts as geometric-dielectric cavity with well-defined photonic density of states, and (3) the epitaxial Fe film hosting the magnonic excitations. Before studying the STO/Fe heterostructure, we start by characterizing separately each of its interacting components, namely the STO cavity and the Fe magnonic medium. 

\subsection{STO photonic cavity}

A simplified analytical expression to describe the eigenmode frequencies of a dielectric box cavity, considering only the material geometry and relative permittivity of the STO substrate, is given by the Fabry-P\'erot cavity mode equation: 
\vspace{10 pt}
\begin{equation}
    \omega_{p}(m,n,p) = \frac{c\pi}{\sqrt{\epsilon_r(T)}} \sqrt{(\frac{m}{L_x})^2 + (\frac{n}{L_y})^2 + (\frac{p}{L_z})^2}
    \label{eq:mode_equation}
\end{equation} 
where \textit{c} is the speed of light in vacuum, $L_x=L_y$ = 10 mm and $L_z$ = 0.5 mm are the dimensions of the STO substrate along the three spatial directions, and $m$, $n$, and $p$ are integers corresponding to the cavity mode indices. The temperature dependence of the relative permittivity $\epsilon_r(T)$ can be described phenomenologically by the Vendik-Zubko model, which allows to incorporate realistic crystal imperfections, as well as the effect of applied electric field \cite{Vendik1997, VendikvsBarrett}, as will be discussed in Section \ref{gating}.

\vspace{10 pt}

\begin{figure*}[h]
    \centering
    \includegraphics[width=\linewidth]{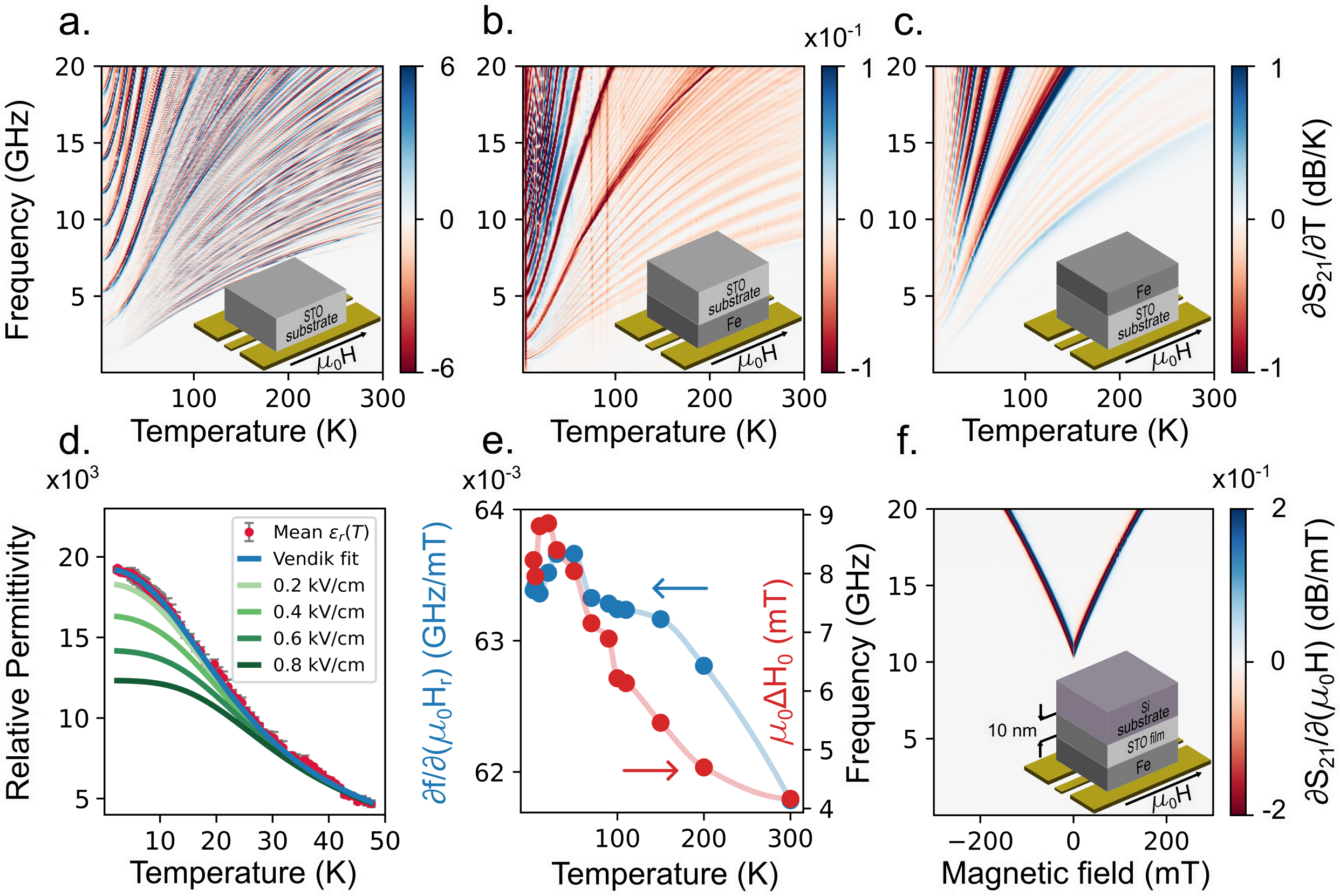}
    \caption{Separate characterization of the dielectric cavity (STO) and the magnonic subsystem (Fe). Top row gives the  $\partial S_{21}(f,T)/\partial T$ spectrum  measured at $\mu_0 \textit{H}$ = 0 mT for (a) The bare STO substrate, (b) The STO/Fe sample in Fe-side excitation and (c) The STO/Fe sample in STO-side excitation. The left axis on panels a-c corresponds to the RF frequency, while the right axis color bar indicates the intensity of the $S_{21}$ partial derivative with respect to temperature.
    Bottom row shows (d) Temperature dependence of the effective dielectric permittivity $\epsilon_r(T)$, extracted from the resonant mode evolution of bare STO, below $T$ = 50 K, using Equation 1. Blue curve shows the fit to the Vendik model at zero electric field, while the green curves show the predicted impact of electric field on the temperature dependence of permittivity. (e) Slope $\partial f / \partial (\mu_0 H_r)$ (blue) and linewidth $\mu_0\Delta H_0$ (red) as a function of temperature, extracted from the ferromagnetic resonance spectra of the reference sample Si (substrate)/STO (10 nm)/Fe (30 nm). An example of one of the spectra is shown for $T=$ 300 K in (f), by plotting the heatmap of $\partial S_{21}$($f,\mu_0$H)/$\partial (\mu_0H)$ as a function of $\mu_0H$ and $f$.}
    \label{fig:DielectricCavity}
\end{figure*}

As a first reference measurement, the S$_{21}$ transmission parameter of the microwave stripline, coupled to a bare STO crystal, was measured as a function of temperature, with the stripline aligned along the STO$\langle110\rangle$ crystalline axis. To highlight the temperature-dependent features in the resonant mode spectrum originating from the STO substrate, the evolution of the partial derivative of $S_{21}$ with respect to temperature is shown in Figure \ref{fig:DielectricCavity}a. The pronounced absorption lines correspond to the expected standing wave cavity modes excited within the STO crystal, described by Equation \ref{eq:mode_equation} and dominated by the thickness mode number $p$, especially at low temperatures. At higher temperatures (above 50 K), both vertical and lateral modes are prominent. By tracking the measured absorption peak frequencies as a function of temperature for $T<$ 50 K, the measurement provides a way to determine the effective microwave permittivity of the STO crystal (Supplementary Figure S3, S4), allowing us to extract the temperature-dependent values of the effective permittivity $\epsilon_r(T)$ (Figure \ref{fig:DielectricCavity}d, red points) and fit this dependence to the Vendik-Zubko parameters (blue curve). The green curves represent the model's predicted impact of electric field application on this permittivity evolution, discussed in Section \ref{gating}. A detailed outline of the quantitative procedure to extract the effective microwave permittivity can be found in Supplementary Section 3.

\vspace{10 pt}

Following the characterization of the bare STO dielectric cavity, the STO/Fe heterostructure was subjected to the same measurement protocol. For this sample, two distinct configurations were characterized: (I) \textit{Fe-side excitation}, with the Fe film facing down, strongly participating in the microwave coupling to the stripline, and (II) \textit{STO-side excitation}, with the Fe film side facing up, separated from the stripline by the STO substrate. In the case of Fe-side excitation, shown in Figure \ref{fig:DielectricCavity}b, the metallic Fe overlayer strongly modifies the dielectric cavity mode spectrum compared to the bare STO substrate. Firstly, it introduces an additional sheet conductance, lowering the amount of RF radiation that leaks effectively into the cavity, due to dissipative microwave shielding currents \cite{Baumgarten26}. Secondly, the metal-coated side imposes a boundary condition with a reflection coefficient distinct from the dielectric-air surface. The difference in reflection properties between the top and bottom cavity walls of the STO substrate causes a reconfiguration of the mode spectrum, supporting additional quarter-wave resonances at low temperatures. These considerations explain the substantially reduced absorption amplitude, as well as the nearly twice as small spacing between the vertical $p$-mode frequencies at the lowest temperatures in Figure \ref{fig:DielectricCavity}b. The metallic boundary condition furthermore limits the amount of higher order lateral modes that survive compared to the bare STO crystal. For the STO-side excitation (Figure \ref{fig:DielectricCavity}c), the sample was cut to 4 x 4 mm$^2$ lateral dimensions in order to displace a large fraction of lateral mode excitation outside the $f$ = 0-20 GHz measurement range, as will be discussed in Section \ref{MPcoupling:coherent}. Since metallic screening is absent when Fe is on the STO side opposite to the one contacting the stripline, the absorption amplitude is much less suppressed. Additionally, comparing Figures \ref{fig:DielectricCavity}b and \ref{fig:DielectricCavity}c, the lateral confinement makes the vertical $p$-modes even more prominent.

\subsection{Fe magnonic medium} \label{subFecharacterization}

To study the isolated microwave response of the Fe magnonic medium, a thin film of STO (10 nm) was deposited on single crystal Si(001) substrate, forming a reference growth template for epitaxial deposition of the metallic Fe film. The limited thickness of the STO film moves its photonic response far above the explored microwave frequency range, so that no magnon-photon interactions should occur below $f=$ 20 GHz. This allows us to study intrinsic magnetic microwave properties of the Fe film as a function of frequency and temperature, while maintaining nearly identical structural properties as the Fe film deposited on the bulk single crystal STO substrate. Figure \ref{fig:DielectricCavity}f shows the FMR resonance spectrum $\partial S_{21}$($f,\mu_0H$)/$\partial(\mu_0H)$ along the Fe$<$100$>$ easy axis direction, measured in the Fe-side excitation configuration at $T=$ 300 K. A sharp $k_m = 0$ Kittel magnon mode can be observed, exhibiting clear magnetic easy axis behavior, with a non-zero resonance field $f_r(\mu_0H=0)$ = 10.4 GHz. For an isolated precessing  ferromagnetic system, the linewidth evolution is captured by the relation \cite{Urban2001, Heinrich2002}: 
\begin{equation}
    \mu_0\Delta H(f_r) =  \mu_0\Delta H_0 + \frac{4\pi\alpha}{\gamma} f_r
    \label{standardFMR}
\end{equation} 
where $\mu_0\Delta H_0$ is the frequency-independent linewidth contribution, $\gamma$ is the gyromagnetic ratio, $\alpha$ is the Gilbert damping parameter and $f_r$ the ferromagnetic resonance frequency. Using a linear fit procedure, the $\mu_0\Delta H_0$-value was extrapolated, using Equation \ref{standardFMR} at each temperature. With decreasing temperature, the average slope $\partial f/\partial (\mu_0H_r)$ (where $H_r = H(f_r)$) and resonance linewidth $\mu_0\Delta H_0$ increase, as shown in Figure \ref{fig:DielectricCavity}e. The evolution of both values is consistent with earlier reported measurements on epitaxial Fe thin films \cite{FeEvolutionofLWandT}. The increase in the FMR slope $\partial f/\partial (\mu_0H_r)$ with decreasing temperature, as well as the decrease in resonance linewidth, are directly associated with the temperature dependence of saturation magnetization and magnetocrystalline anisotropy energy. The results are consistent with the SQUID-VSM characterization, shown in Supplementary Figure S2c. The combination of low intrinsic damping ($\alpha_{T = 300 K} = 1.8 \times10^{-3}$ ) and high $\mu_0H_k$ positions the epitaxial Fe film as a high-quality magnonic medium. 

\section{Magnon-photon coupling in the STO/Fe magnonic cavity material} \label{MPcoupling}

Having established a separate characterization of the STO cavity subsystem and the Fe magnonic medium, we then investigated the field-dependent microwave transmission $\partial S_{21}$($f,\mu_0H$)/$\partial (\mu_0H)$ of the combined STO/Fe magnonic cavity system, measured at different temperatures. Firstly, we correlate the temperature-dependent cavity mode frequencies, identified earlier in Figure \ref{fig:DielectricCavity}b,c, to modulations in the ferromagnetic resonance spectrum at corresponding temperatures. This initial analysis provides direct evidence for controlled magnon-photon coupling in the STO/Fe system. Secondly, we analyze the hybridized magnon-photon spectra to extract the relevant coupling parameters for a phenomenological model of the STO/Fe system, based on a coupled-mode Hamiltonian. An important aspect concerning circuit integration of magnon-photon coupling, is that the observed microwave response strongly depends on the excitation configuration. In the following, we perform our analysis on the experimentally determined resonance spectra for Fe- and STO-side excitation, dominated by dissipative and coherent magnon-photon coupling, respectively. 

\vspace{10 pt}

\subsection{Fe-side excitation: dissipative interaction} \label{MPcoupling:dissipative}

\begin{figure*}[h!]
    \centering
    \includegraphics[width=0.9\linewidth]{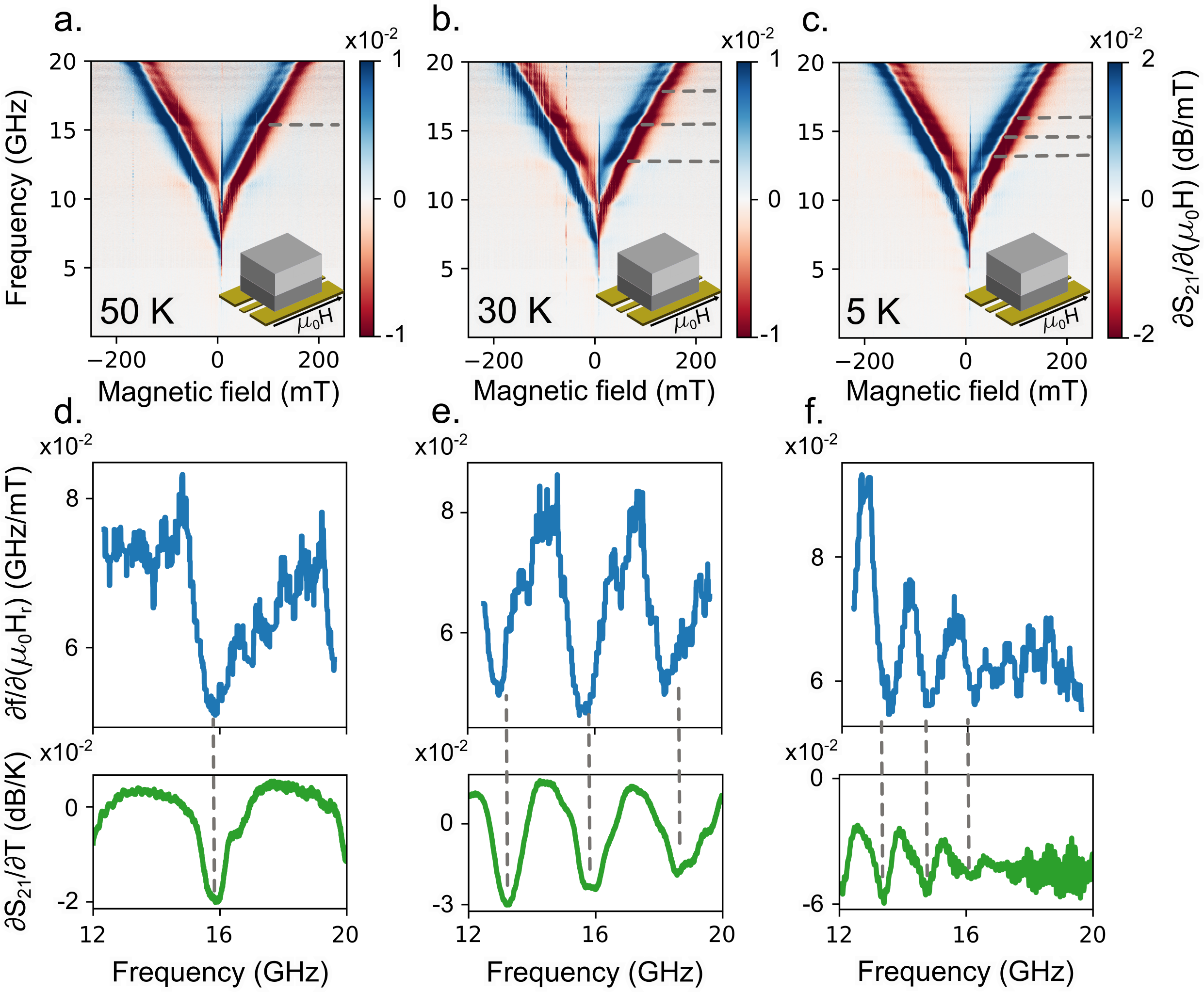}
    \caption{(a-c) Ferromagnetic resonance characterization of the STO (gray)/Fe (dark gray) system at different temperature values ($T$ = 50 K, 30 K and 5 K), in the form of $\partial S_{21}$($f,\mu_0H$)/$\partial (\mu_0H)$ heatmaps. The left axis on panels a-c corresponds to the RF frequency, while the right axis color bar indicates the intensity of the $S_{21}$ partial derivative with respect to magnetic field. (d-f) Correlation of the FMR line slope $\partial f/\partial (\mu_0 H_r)$ (blue) with the cavity resonance spectrum $\partial S_{21}(f,T)/\partial T$ (green) from Figure \ref{fig:DielectricCavity}b, at the corresponding temperatures.}
    \label{fig:FMR}
\end{figure*}

Excitation by the microwave stripline from the Fe film-side causes both the magnonic medium (Fe) and the microwave active substrate (STO) to interact with the coplanar waveguide, forming an \textit{open cavity} system \cite{Harder21, LiHybridMagnonics}. Figure \ref{fig:FMR}a-c shows the $\partial S_{21}$($f,\mu_0H$)/$\partial (\mu_0H)$ heatmaps at $T = $ 50 K, 30 K and 5 K, respectively, measured along the Fe magnetocrystalline easy axis, for Fe-side excitation. Showing the derivative of the transmission parameter with respect to magnetic field removes the field-independent STO resonance modes and the background induced by the RF-wiring and the sample holder. This allows to clearly identify the ferromagnetic resonance response and assess how it is affected by the magnon-photon interaction with the STO substrate. The heatmaps for all measured temperatures, as well the measurements along the Fe $<$110$>$ magnetocrystalline hard axis, can be found in Supplementary Figures S5 and S6, respectively. While exhibiting a similar Kittel-like response to the Fe reference sample discussed in Subsection \ref{subFecharacterization}, a first important observation is that the linewidth of these resonance curves ($\mu_0\Delta H_0 \approx$ 34 mT) is substantially larger than for the reference sample ($\mu_0\Delta H_0 \approx$ 4 mT). Given that the measurement conditions except for the substrate are identical, this can be attributed predominantly to an additional radiative magnon decay rate induced by the STO resonant cavity \cite{Yao2019RadiativeDamping}.

\vspace{10 pt}

In order to make the presence and origin of magnon-photon coupling more apparent, the field-dependent microwave response at each temperature was correlated with the corresponding dielectric cavity spectrum of the STO/Fe system, characterized in the previous section, as a function of temperature. Figure \ref{fig:FMR}d-f shows a correlation between the resonance dips of the STO cavity (green), extracted from Figure \ref{fig:DielectricCavity}b, to the slope of the ferromagnetic resonance frequency as a function of magnetic field (blue) at corresponding temperatures. The grey dashed lines correlate the STO cavity modes to the pronounced slope suppressions in the FMR resonance line. Furthermore, the observed slope modulations are accompanied by local intensity enhancements in the hybrid mode spectrum near the cavity mode frequencies. Both of these observations are a result of the dissipative magnon-photon interaction, as will be discussed in Subsection \ref{MPCoupling:model} and Supplementary Subsection S4.3 (Figure S9-S12).

\vspace{10 pt}

Summarizing these first observations of magnon-photon coupling, the Fe-side excitation configuration induces significant FMR linewidth broadening, as well as pronounced local changes in the FMR slope. The frequency dependence of these local modulations correlates directly with the earlier identified STO cavity $p$-mode frequencies, indicative of an open cavity-type dissipative coupling in the STO/Fe system. While dissipative coupling is present in any practical small-scale implementation, changing the excitation configuration can strongly enhance coherent coupling, as will be demonstrated in the next section.

\subsection{STO-side excitation: coherent interaction} \label{MPcoupling:coherent}

To minimize direct interaction of the ferromagnetic Fe film with the microwave stripline, the sample was mounted with the STO-side contacting the coplanar waveguide. Due to the absence of microwave screening by the metallic Fe film, STO-side excitation allows for more efficient cavity mode excitation in the STO substrate. Additionally, it removes to a large extent the cooperative damping channel, since the magnon response in Fe interacts predominantly directly with STO cavity modes, rather than with the microwave stripline. This configuration allows to more effectively study coherent magnon-photon coupling in the STO/Fe system. Given that the suppression of lateral cavity modes improves the prominence of thickness-defined and larger spaced $p$-modes (Figure \ref{fig:DielectricCavity}c), the sample was cut to a size of 4 x 4 mm$^2$, more closely matching the stripline dimensions.

\vspace{10 pt}

\begin{figure*}[h!]
    \centering
    \includegraphics[width=\linewidth]{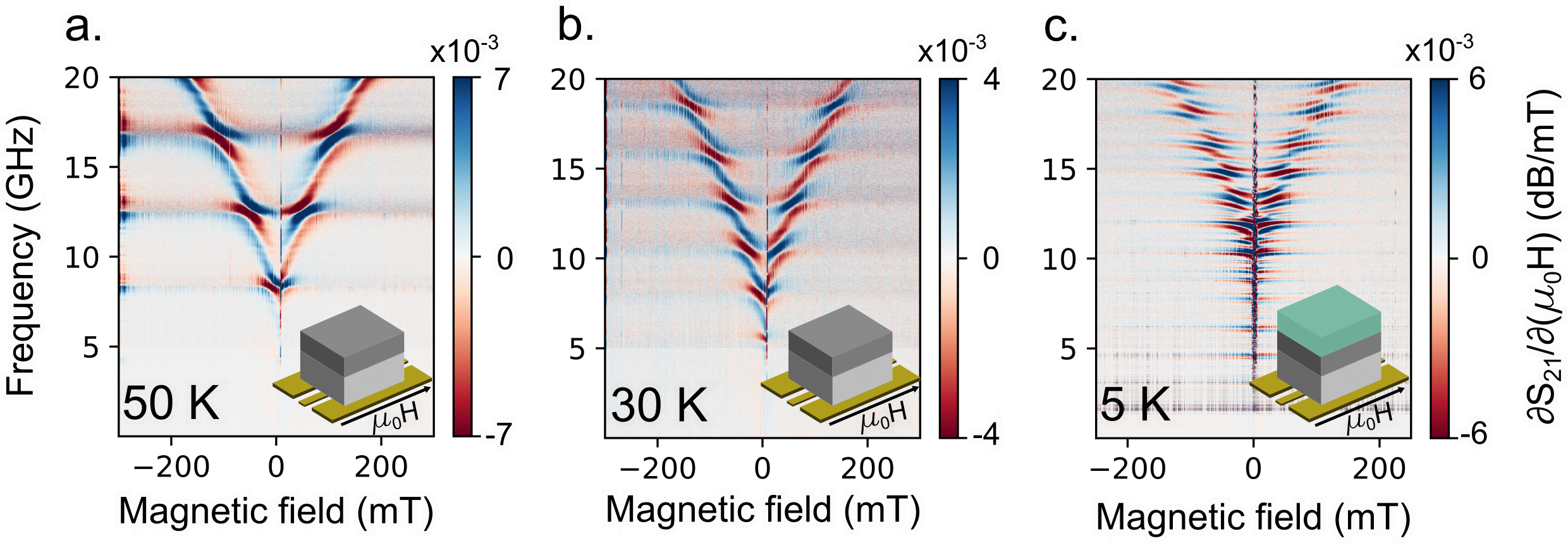}
    \caption{(a-b) Ferromagnetic resonance characterization of the STO (gray)/Fe (dark gray) system in STO-side excitation, at different temperature values ($T$ = 50 K and 30 K), in the form of $\partial S_{21}$($f,\mu_0H$)/$\partial (\mu_0H)$ heatmaps. (c) STO (gray)/Fe (dark gray)/Nb (green) system excited in the same manner at $T$ = 5 K. Low-temperature modes are maintained prominently compared to the STO/Fe sample, due to the presence of the thicker Nb layer, which `sharpens' the hybridized mode spectrum. The left axis on all panels corresponds to the RF-frequency, while the right axis color bar indicates the intensity of the $S_{21}$ partial derivative with respect to magnetic field.}
   
    \label{fig:FMRflipped}
\end{figure*}

Figures \ref{fig:FMRflipped}a,b show the $\partial S_{21}$($f,\mu_0H$)/$\partial (\mu_0H)$ heatmaps for temperatures $T = $ 50 K and 30 K in STO-side excitation. Measurements for other temperatures can be found in Supplementary Figure S7. It is immediately clear that this configuration causes more effective magnon-photon mode hybridization, exhibiting avoided crossing features with enhanced intensity in the field-dependent microwave response. As shown in Figure \ref{fig:DielectricCavity}c, Supplementary Figures S7a,b and S11c,f, the intensity of the hybridized modes strongly fades below $T=$ 20 K. In this regime, the modes are densely spaced in frequency, so that their coupling and damping rates start to exceed the frequency separation of the cavity modes. As a result, the mode intensities spread out and diminish, consistent with our model in Supplementary Figure S11f. In fact, the intensity reduction provides indirect evidence that the excitation of ferromagnetic precession in the Fe film for this configuration is mediated by the STO cavity resonances, rather than by direct interaction with the $\mu_0H_{RF}$-component generated by the microwave stripline, which would exhibit a less prominent temperature dependence. 

\vspace{10 pt}

To combat the observed `fading' of the hybridized mode spectrum at the lowest temperatures, we deposited a 85 nm superconducting Nb film ($T_c \approx$ 8.7 K, Supplementary Figure S2d) on top of the Fe layer. The addition of a superconductor overlayer should drastically reduce the effective resistance of the metallic `mirror' on the STO surface, enhancing the cavity quality factor. Figure \ref{fig:FMRflipped}c shows that the mode spectrum of the STO/Fe/Nb sample at $T=$ 5 K is noticably sharper and more intense compared to the STO/Fe sample at the lowest temperatures (Supplementary Figure S8 compared to S7a,b). It is important to stress that the mode sharpening and intensity enhancement compared to STO/Fe are not directly related to superconductivity, as shown in Supplementary Figure S8a (measured at $T=$ 10 K $>T_c$). The thicker metallic Nb layer already lowers the sheet resistance significantly. Interestingly, the mode spectrum in the superconducting state seemingly supports more FMR resonance branches than only the $k_m$=0 magnon mode (Supplementary Figure S8b,c), as reported in earlier works \cite{GolovchanskiySFHybrid, AFMGolovchanskiy, golovchanskiy2019interplay}. While a detailed study of the impact of superconductivity is outside the scope of the current work, it could provide an important route to additional control of the magnonic response and cavity quality factor through material design.

\subsection{Phenomenological model: coupled mode Hamiltonian}\label{MPCoupling:model}

In the previous section, we demonstrated effective hybridization between the STO dielectric cavity modes and the Fe ferromagnetic resonance Kittel mode. In this section, we obtain quantitative estimates for the dissipative and coherent magnon-photon strength by analysis of the hybridized mode spectra. Based on these results, we establish a simplified phenomenological model to describe the full system response, using the effective coupled-mode Hamiltonian \cite{WangMing_Dissipative_cavitymagnonics}:

\begin{equation}
\mathcal{H}_{eff}(\mu_0 H) =
\begin{pmatrix}
\omega_p(k_p)-\frac{i\kappa_p}{2} & .. & g+i\gamma_{diss} \\ & .. & & \\
g-i\gamma_{diss}& ..  & \omega_m(\mu_0 H;k_m=0)-\frac{i\kappa_m}{2}
\end{pmatrix}
\label{hamiltonian}
\end{equation}

\vspace{10 pt}

The diagonal elements of this coupling matrix represent the STO photonic cavity ($\omega_p$)- and Fe magnonic mode ($\omega_m$) frequencies, with radiative photon- and magnon decay rates $\kappa_p$ and $\kappa_m$, respectively. On the off-diagonal elements, the coupling term $g+i\gamma_{diss}$ describes the coherent-dissipative interaction between the photonic and magnonic subsystems, mediated by direct magnon-photon coupling and cooperative damping in the electromagnetic environment of the open cavity \cite{Yang24}. The microwave transmission parameter $S_{21}(H,f)$ is then computed using the complex eigenfrequencies of Equation \ref{hamiltonian}.

\vspace{10 pt}

\begin{figure*}[h!]
    \centering
    \includegraphics[width=\linewidth]{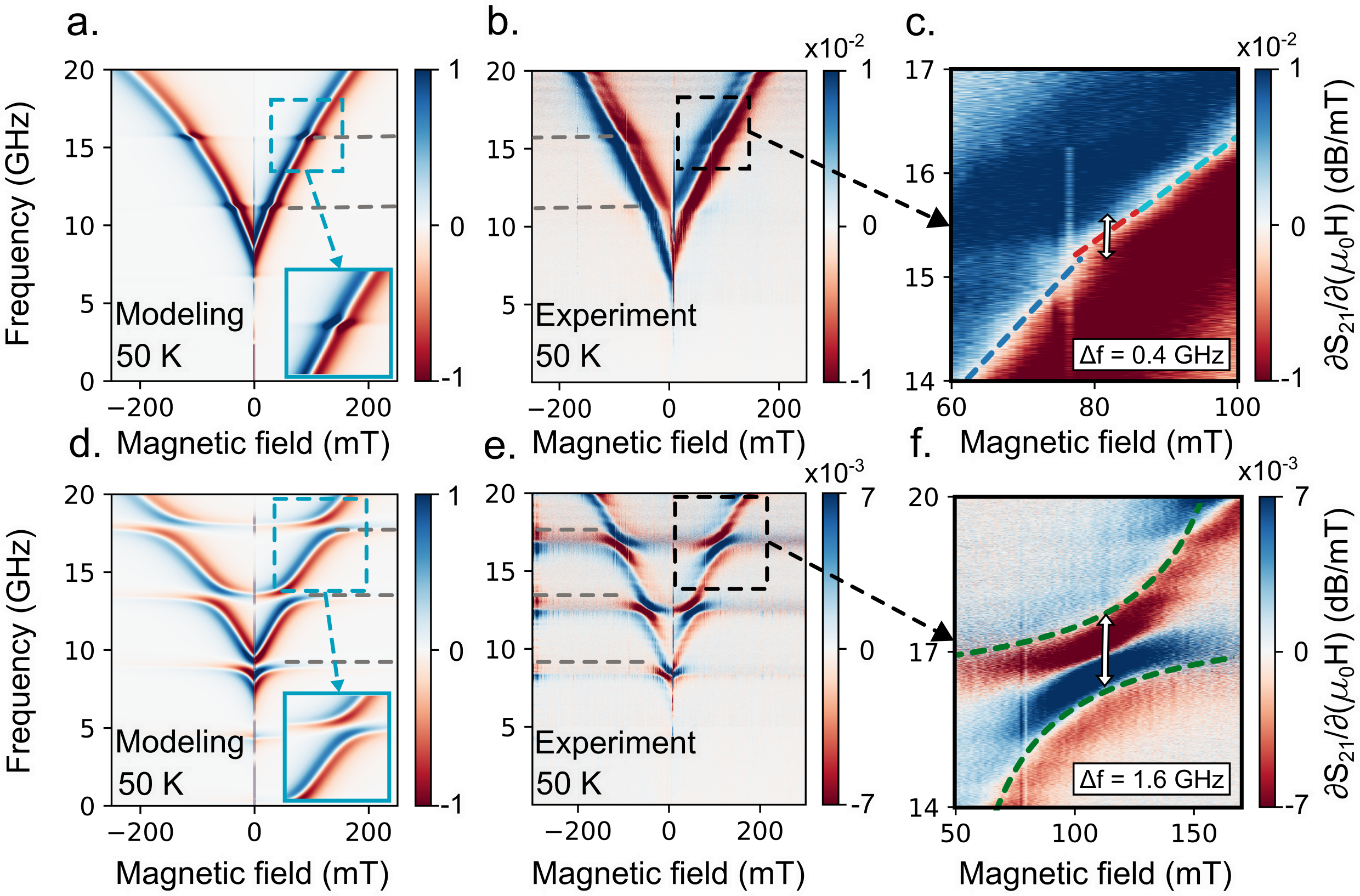}
     \caption{Phenomenological modeling, shown here for $T$ = 50 K. Simulated (a) and measured (b) mode spectra for Fe-side excitation. Grey dashed lines show the correspondence of the mode frequencies between model and experiment. (c) Quantitative estimate of the dissipative coupling strength $\gamma_{diss} \approx \frac{\Delta f}{2}$, zoomed in around the $f$ = 15.4 GHz mode. The colored dashed lines provide a guide to the eye for the local slope modification of the $k_m=$ 0 magnon mode. Simulated (d) and  measured (e) mode spectrum for STO-side excitation. (f) Quantitative estimate of the coherent coupling strength $g \approx \frac{\Delta f}{2}$, zoomed in around the $f=$ 16.5 GHz mode. The green dashed lines provide a guide to the eye for the hybrid mode splitting.}

    \label{fig:parameter_extraction}
\end{figure*}

Based on this formalism, we simulate the $S_{21}(f,\mu_0H)$ transmission spectra and plot the corresponding heatmaps of $\partial S_{21}$($f,\mu_0$H)/$\partial (\mu_0H)$. Figure \ref{fig:parameter_extraction}a,d show the simulated $\partial S_{21}$($f,\mu_0$H)/$\partial (\mu_0H)$ heatmaps by the Hamiltonian model at $T$ = 50 K, while the corresponding measured data are shown in Figure \ref{fig:parameter_extraction}b,e, respectively. Quantifying the frequency range $\Delta f$ of the slope suppression (dissipative coupling) and mode splitting (coherent coupling), we obtain quantitative parameter estimates for $\gamma_{diss} \approx 0.2$ GHz and $g \approx 0.8$ GHz, equal to half the frequency range $\Delta f/2$, as shown in Fig \ref{fig:parameter_extraction}c,f. This procedure is further discussed in Supplementary Section 4.3 and Figure S9. The damping parameters $\kappa_p = 0.6$ GHz and $\kappa_m = 2$ GHz were estimated from the average cavity mode full width at half maximum (FWHM) and from the frequency-independent linewidth $\mu_0\Delta H_0 \times \frac{\partial f}{\partial(\mu_0H_r)}$ of the bulk STO/Fe sample, respectively. Comparison of the experimental data with the modeling for Fe-side and STO-side excitation at multiple temperatures is shown in Supplementary Figures S10 and S11, respectively.  

\vspace{10 pt}

While each off-diagonal element in the coupling Hamiltonian should in principle be identified separately as function of frequency and temperature for an accurate description of coherent-dissipative coupling for all individual modes, we use a simplified approach, quantifying an average representative coupling strength. We extracted this value for the $T =$ 50 K measurement, and use it as an effective coupling term representative for all temperatures. The hybridized modes are well-separated for this temperature, avoiding spectral overlap. While this approach overestimates the coupling strength at low temperatures and high frequencies, it enables a satisfactory reproduction of the most important physical features. The magnon-photon eigenmodes, resulting from Equation \ref{hamiltonian} are then given by the expression \cite{energy_splitting_10}:

\begin{equation}
        \omega_{m-p} = \frac{\omega_m +\omega_p}{2} -i\frac{\kappa_m +\kappa_p}{4} \pm \sqrt{(\frac{\omega_m -\omega_p}{2} -i\frac{\kappa_m +\kappa_p}{4})^2+(g-i\gamma_{diss,mp})^2}
    \label{eq:eigenfreqs}
\end{equation}

\vspace{10 pt}

For the Fe-side excitation, the coherent coupling term $g$ is negligible and set to 0 GHz for simplicity. The dissipative term $\gamma_{diss}$ is set to 0.2 GHz. Equation \ref{eq:eigenfreqs} shows that dissipative coupling causes a splitting in the imaginary part of the eigenmode frequencies, leading to enhanced intensity of the Fe FMR derivative above/below the STO cavity resonance for the inner/outer branch. This is shown for the model in Figure \ref{fig:parameter_extraction}a, and experimentally in Figure \ref{fig:parameter_extraction}b,c. Supplementary Figure S12 correlates the maxima in the experimental and simulated absolute intensities $|\partial S_{21}$($f,\mu_0$H)/$\partial (\mu_0H)|$ to the STO cavity modes. This approach is complementary to the earlier identification of local slope modulations, which was discussed in Subsection \ref{MPcoupling:dissipative}. For the STO-side excitation, $\gamma_{diss}$ is set to 0, and $g$ to 0.8 GHz. This coupling causes a splitting in the real part of the hybrid eigenmode frequencies, giving rise to the characteristic avoided crossings observed in Figure \ref{fig:parameter_extraction}d (model) and e,f (experiment). As for the case of dissipative coupling, the intensities in both the simulated and experimental FMR resonance spectra are maximal just above and below the cavity mode frequencies. The dashed lines in Figures  \ref{fig:parameter_extraction}c,f provide a guide to the eye for the mode splitting.

\vspace{10 pt}

Our simplified model captures the most prominent effects of magnon-photon coupling on the microwave transmission spectrum, and its evolution with temperature. Furthermore, it highlights the inherent tradeoff between practical microwave excitation and the coherence of interaction when using magnonic cavity materials. 

\newpage
\vspace{10 pt}

\section{Voltage control of the magnon-photon coupling}\label{gating}

As discussed in the introduction, the dielectric permittivity of STO is strongly affected by an externally applied electric field, particularly in the temperature regime below $T$ = 50 K, as shown in Figure \ref{fig:DielectricCavity}d \cite{Muller79, Vendik1997, Neville72}.  Application of a gate voltage is hence expected to modify the STO dielectric permittivity value $\epsilon_r(T,V_G)$ and thereby shift the cavity mode frequencies according to Equation \ref{eq:mode_equation}. Such intrinsic, nonlinear voltage-tunability provides a direct means to electrically control the hybridized magnon-photon mode spectrum, a major advantage with regards to scalable integration and local tunability. The effect of electric field is maximal at the lowest temperatures, which are the most relevant for integration into superconducting (quantum) electronics. Therefore, we choose to focus on the deep cryogenic regime, namely at a temperature of $T$ = 4 K. Due to practical considerations regarding the application of gate voltage in the setup (RF-field screening by the bottom electrode), only the Fe-side excitation configuration was measured. However, the gate effect should be equally applicable to both coherent and dissipative coupling. 

\vspace{10 pt}

To enable the application of a bottom gate voltage, a Ti(8 nm)/Nb(200 nm) metallic back electrode was deposited by MBE on the (non-polished) backside of the STO substrate. The magnetic field-dependent microwave response was measured for positive and negative gate voltages between 0 V and 40 V ($E$ = 0.8 kV/cm), in steps of 10 V ($\Delta E$ = 0.2 kV/cm). Characterization of the $S_{21}(f,H)$ parameter at all gate voltages, showing the STO resonance lines and their gate-induced shifts, can be found in Supplementary Figures S13, S14 and S15. Figure \ref{fig:gating}a shows that the presence of the back contact significantly modifies the observed resonance spectrum, by altering the boundary conditions on the STO back side and the effective sample microwave impedance, establishing multiple microwave reflections in the sample \cite{Understanding_the_Line_Shape_of_Ferromagnetic_Resonance_Spectrum_in_a_Flip-Chip_Measurement}. The FMR response then becomes asymmetric: the inner branch becoming more intense and narrow, while the outer branch becomes more spread out. The induced asymmetry causes the intensity modulations in the outer branch to become visually more pronounced. As established in Subsections \ref{MPcoupling:dissipative}, \ref{MPCoupling:model} and Supplementary Subsection 4.3, the positions of these intensity modulations correlate to the STO cavity mode frequencies.

\vspace{10 pt}

Figure \ref{fig:gating}b shows a trace of the transmission parameter derivative as function of magnetic field for a fixed reference frequency $f_{ref}$ = 14 GHz (green dashed line in Figure \ref{fig:gating}a), under the application of different back-gate voltages. It is clear that the intensity of both the inner- and outer FMR branch change substantially with gate voltage shown in Figure \ref{fig:gating}c in dark blue and red, respectively. The opposite evolution for inner-and outer branch can be explained by the observations discussed in Subsection \ref{MPCoupling:model} and Supplementary Section 4.3: the intensity of the peak and dip in the FMR derivative reside above and below the cavity mode frequency, respectively. Since the field-induced mode shifts at $\Delta E = 0.2$ kV/cm steps are of similar magnitude as the frequency spacing between the cavity modes, the intensity maximum on the outer branch (dip) typically moves towards $f_{ref}$ when the maximum on the inner branch (peak) passes it, moving upwards. In this sense, the dynamic magnetic response at a fixed driving frequency and static magnetic field is strongly tunable by gate voltage. It should be noted that this is not due to a modification of intrinsic magnetic properties of the Fe film, but by a electrical reconfiguration of its magnonic response when hybridizing with the photonic STO cavity. 

\vspace{10 pt}

According to the Vendik-Zubko model discussed earlier, the application of an electric field reduces the dielectric permittivity of STO, shifting the cavity modes towards higher frequencies while their relative spacing increases \cite{Vendik1997}. Figure \ref{fig:gating}d shows the difference between ferromagnetic resonance spectra at gate voltages $V_G$ = 30 V and 0 V. The alternating positive-negative (blue-red) sign of the intensity difference along the FMR resonance curve indicates an effective shifting of the hybrid modes under the applied electric field. To capture this effect quantitatively, we add the phenomenological electric-field dependence to the $\epsilon_r$-value in Equation \ref{eq:mode_equation}, which we then use to compute the diagonal elements $\omega_p(k)$ of the coupled-mode Hamiltonian. Keeping the off-diagonal coupling terms constant, we can reproduce to satisfactory extent the electric-field dependence of the mode spectrum, shown in Supplementary Figure S13 for multiple gate voltages with respect to $V_G$ = 0 V. Focusing on the frequency window of interest ($f$ = 11.5 GHz - 16 GHz) demarcated by the dashed black lines in Figure \ref{fig:gating}a, we correlate the mode evolution of the metal-coated STO cavity (Figure \ref{fig:gating}e) with that of the hybrid magnon-photon modes of the STO/Fe system under electric field application (Figure \ref{fig:gating}f). Figure \ref{fig:gating}e shows the $S_{21}$ transmission parameter, measured at a fixed magnetic field of $\mu_0H$ = 250 mT as a function of frequency and gate voltage. The average signal over all gate voltages was subtracted as a baseline, yielding $\Delta S_{21}(f,V_G)_{\mu_0H=250~mT} = [S_{21}(f,V_G) - S_{21}(f)_{avg}]_{\mu_0H=250~mT}$. Since the applied magnetic field lies far outside the field region for ferromagnetic resonance, this intensity map is representative of the gate-voltage evolution of the non-hybridized STO cavity modes.

\begin{figure*}[h!]
    \centering
    \includegraphics[width=0.9\linewidth]{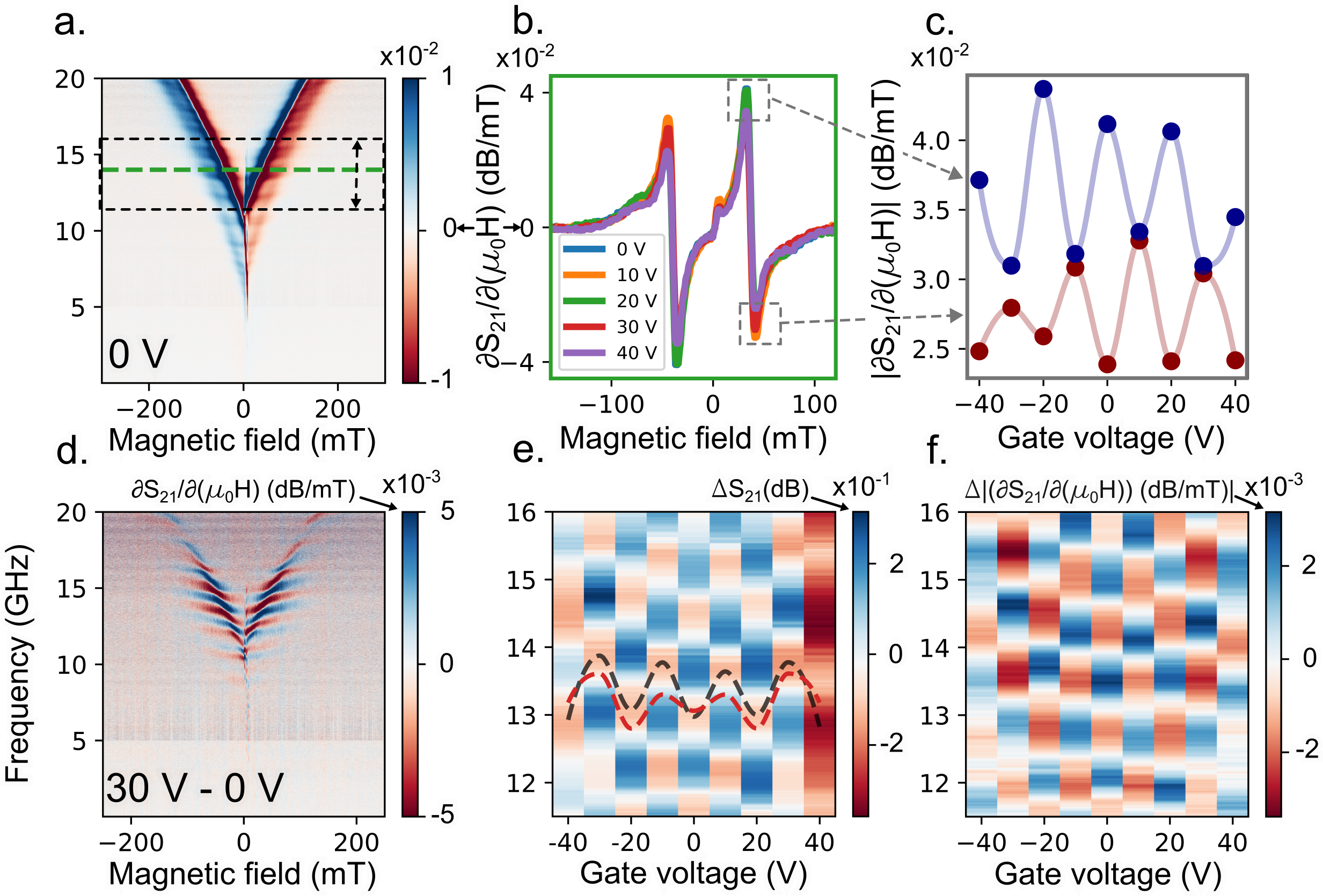}
     \caption{(a) FMR derivative heatmap measured at $T = 4$ K and $V_g = 0$ V. (b) FMR cross-sections taken along the green line of panel a ($f_{ref}$ = 14 GHz), for different back-gate voltages. (c) Intensity along inner (top) and outer (bottom) FMR derivative branch as a function of back-gate voltage. The anti-phase evolution follows naturally from the mode intensity shifts (d) Difference between the heatmaps measured at $30$ V and $0$ V, at temperature $T = 4$ K. (e) Raw FMR data measured at $250$ mT, far outside of the FMR resonance condition. The average transmission over all gate voltages was subtracted as baseline, emphasizing the voltage-induced shifts of the STO resonances. The black dashed line indicates the evolution of the STO resonances in the frequency range from $12.5$ GHz to $14$ GHz. The red dashed line indicates the mode evolution predicted by the Vendik–Zubko model. (f) FMR dip depth (absolute value) in the outer branch, as a function of gate voltage for different frequencies. The heatmap shows the data after baseline subtraction of the average over all gate voltages, to remove the background signal. The hybrid STO/Fe mode spectrum correlates to the STO cavity resonances in panel e.}
    \label{fig:gating}
\end{figure*}

From an application perspective, it is highly relevant to quantify to what extent the electric field can shift the mode frequencies near a chosen reference $f_r$. To this end, the black dashed line in Figure \ref{fig:gating}e highlights the STO mode evolution as a function of gate voltage around $f_r$ = 13.5 GHz, by connecting the nearest $S_{21}$ transmission dips for each gate voltage. The red dashed line shows the same evolution, predicted by the Vendik-Zubko model. The model tends to underestimate the mode shifts at $\pm$ 10 V, possibly because it does not capture the asymmetry of the resonance spectrum, and assumes strictly half-wave resonances with idealized, Lorentzian shape. Nevertheless, the model reproduces the experimental trend of the gate voltage evolution of the STO cavity modes, which can be electrically shifted by nearly $\frac{\Delta f}{\Delta V_G} \approx$ 100 MHz/V. 

\vspace{10 pt}

Having established the impact of electric field on the cavity modes outside the ferromagnetic resonance condition, we then perform the same analysis for the hybrid magnon-photon modes. To this end, we extract the intensity of the dip in the outer branch of the derivative FMR curve (the dip depth evolution shown in red in panel c for $f$ = 14 GHz) for each frequency within 11.5-16 GHz. Again, the average signal over all gate voltages was subtracted as a baseline, yielding $\Delta |\partial S_{21}(f,V_G)/\partial(\mu_0H)| = |\partial S_{21}(f,V_G)/\partial(\mu_0H)| - |(\partial S_{21}(f)/\partial(\mu_0H))_{avg}|$. The result is shown in Figure \ref{fig:gating}f. The minima (blue) and maxima (red) in the dip intensity in the $(f,V_G)$-plane represent the hybridized modes, and correlate very well to the STO mode positions, identified in Figure \ref{fig:gating}e. These intensity modulations evolve in the exact same manner as the gate-induced STO cavity mode movement. It should be noted that, while the $V_G$-axis is discretized in steps of 10 V, the $f$-axis is highly resolved, to steps of $\Delta f=$ 2 MHz. The correspondence between the mode frequency evolutions for the STO transmission dips and the hybridized mode intensities unambiguously proves that the gate-tunability of the magnon-photon modes originates from the electric field-dependence of the STO dielectric permittivity.

\section{Conclusion}

In this work, we realized a magnonic cavity material consisting of an epitaxial STO/Fe(001) heterostructure, for which we demonstrated and quantified both dissipative ($\gamma_{diss}\approx$ 0.2 GHz at $T=$ 50 K) and coherent magnon-photon coupling ($g\approx$ 0.8 GHz at $T=$ 50 K), and achieved gate-tunable mode hybridization.

\vspace{10 pt}

Starting with a separate characterization of the dielectric cavity- and magnonic subsystems, we then studied the microwave response of the hybrid STO/Fe magnonic cavity heterostructure. During this process, the sample was measured in two distinct configurations: with the Fe film facing the stripline (\textit{Fe-side excitation}), and with the STO-side on the stripline (\textit{STO-side} excitation). For Fe-side excitation, both the magnonic and the cavity subsystem interacted with the microwave stripline, forming an \textit{open cavity} system that exhibited predominantly dissipative interaction. The dissipative interaction could be suppressed by configuring the system in STO-side excitation, achieving substantial coherent coupling between the STO cavity and the Fe ferromagnetic resonance in a wide temperature range. A phenomenological coupled-mode Hamiltonian was used for quantitative description of magnon-photon coupling in both configurations. As a key functionality of the hybrid material system, we demonstrated electrical control of the magnon-photon coupling for the open cavity configuration. The application of a gate voltage across the STO substrate strongly modified its dielectric permittivity, redistributing the cavity mode frequencies and by extension the hybrid magnon-photon modes. We observed consistent modifications of the ferromagnetic resonance spectrum, demonstrating that the magnonic response of the Fe film can be electrically reconfigured, through the dielectric response of STO to an applied electric field. By including the expected electric-field dependence of the STO relative permittivity value into the phenomenological model, we successfully captured the control of cavity- and hybrid modes by means of voltage application.

\vspace{10 pt}

Our experimental results establish the STO/Fe heterostructure as a material platform with substantial, gate-tunable magnon-photon coupling. The intrinsic cavity functionality is highly beneficial for the effective integration of nanoscale metallic magnonic circuits, leveraging the combination of appropriate circuit design with \textit{in-materia}-defined and controlled magnon-photon hybridization. The demonstrated gate-tunability offers an important advancement towards magnetic field- or current-free circuit operation. Building further on the physical mechanisms studied in this work, optimizations of the material stack, such as changing the ferromagnetic thickness, including additional (non-)magnetic layers, STO substrate polishing and thickness reductions could further improve or extend the system functionality, depending on the specific application (minimizing optical or magnetic losses, tuning spin-wave dispersion or magnetic anisotropies, etc.). As such, our work presents a promising route to address the scalability and local tunability challenges of state-of-the art cavity magnonics, using the material-defined, gate-tunable intrinsic magnonic cavity as a complementary approach to circuit design, extending power-efficiency and integration possibilities, specifically in the cryogenic temperature regime of superconducting- and quantum electronics.

\clearpage \newpage

\section{Experimental methods}

\subsection{MBE deposition}
STO single crystalline substrates (Mateck, GmbH) were heated to 300$^\circ$C in the ultra-high vacuum chamber of the Molecular Beam Epitaxy (MBE) system. The temperature was then lowered to 200 $^\circ$C, after which a combined approach of Ti-sublimation and liquid nitrogen cryo-pumping reduced the base pressure of the chamber to $p \approx$ 7$\cdot$10$^{-10}$ mbar. The epitaxial Fe film was deposited by electron gun evaporation at a rate of 0.2 \AA/s, and post-annealed for 20 minutes at 360$^\circ$C to improve the film crystallinity. After letting the sample cool back natively to room temperature, a 6 nm Ti capping layer was deposited on top of the Fe film to protect it from oxidation in ambient environment. For the Fe/Nb sample discussed in Subsection \ref{MPcoupling:coherent}, the Fe sample was transported in the UHV line after deposition, to a second MBE deposition chamber. The Nb deposition was performed there by e-gun evaporation at 0.8 \AA/s. Chamber base pressure was $p \approx$ 8$\cdot$ 10$^{-10}$ mbar. 

\subsection{Structural characterization}

The pole figure measurement of the Fe film on bulk STO substrate (Supplementary Figure S1c) was performed in a Malvern Panalytical X'Pert Pro Diffractometer. The sample was mounted with the STO $<$100$>$-axis aligned to the $\phi$ = 0 azimuthal angle. $2\theta = 44.6 ^\circ$ angle was aligned to the the Fe (011) reflection. The $\chi$ tilt angle of the sample was varied between 20$^\circ$ and 70$^\circ$ in steps of 1$^\circ$, while at each $\chi$-step, the sample was rotated around the surface normal by $\phi = 0 \rightarrow 360^\circ$ in steps of 1$^\circ$. The 4-fold symmetry of the maximal diffraction intensities, appearing at multiples of $45^\circ$ shows the epitaxial relation of Fe with respect to the STO substrate.

\subsection{VSM magnetometry}

The Fe film on bulk STO substrate was measured by SQUID-VSM magnetometry in a Quantum Design MPMS3 system (Supplementary Figure S2). M($\mu_0$H)-measurements were performed in $\mu_0$H = $\pm$ 500 mT range. The M(T) measurements for the STO/Fe sample were done for continuous temperature sweep, cooling down from 300 K to 10 K at $dT/dt$ = 5 K/min. The $T_c$ characterization of Nb was done using a Zero field cooled-heating, Field cooled-cooling measurement between 2 K and 20 K. The VSM oscillation amplitude was 5 mm, and the averaging time 2 seconds for all measurements. Magnetization was obtained by dividing the measured magnetic moment by the estimated Fe sample volume ($V_{Fe}$ = (45 $\pm$ 6) $\cdot$ 10 $^{-8}$ m$^3$) after cleaving.

\subsection{Microwave characterization}

All studied samples were mounted with Kapton tape on a coplanar waveguide (central conductor width 0.6 mm, trench width 0.3 mm), defined on the PCB of the cryo-FMR option in a Quantum Design PPMS Dynacool cryostat. Reported magnetic fields were all applied in the sample plane, along the central axis of the waveguide. The microwave transmission parameter $S_{21}$ was measured using a Keysight Vector Network Analyzer (VNA). Input power for all measurements was 0 dBm, and the frequency sweep range 9 kHz-20 GHz. For gate voltage application in Fe-side excitation (Section \ref{gating}), a thin Au wire was connected on the Fe side and back-electrode side by conductive silver paste. These wires were then connected by silver paste to separate voltage pins on the PCB, which were accessed by the external cabling. No leakage currents were detected exceeding 0.8 nA at any point during the measurements. To limit the impact from dissipation in the Ag-paint contacts, they were placed as far as possible (about 10 mm) from the CPW excitation region. For STO-side excitation of coherent magnon-photon coupling, the requirement of a bottom electrode strongly limits the ability to effectively excite resonant modes in the STO cavity, due to microwave screening. Therefore, no gate experiments were performed in the STO-side excitation. 

\medskip
\newpage
\textbf{Acknowledgements} \par 

The authors would like to acknowledge the contribution of
Julia Baumgarten in the analysis of the STO cavity response. This work is supported
by Research Foundation Flanders (FWO) grant number 11A3V25N (S.R.) and 11BA827N (E.G.),
and under the grant Weave G0D7723N, the Fonds de la Recherche Scientifique - FNRS
under the grant Weave -PDR T.0208.23 and CDR J.0199.25. This work was carried
out during a sabbatical stay of A.V.S at KU Leuven, funded by a scientific mission of
the Fonds de la Recherche Scientifique – F.R.S.-FNRS (Belgium). A.V.S. thanks the
University of Li`ege for granting this research leave, and KU Leuven for its hospitality.
J.V. and K.T. acknowledge the KU Leuven C1-program grant, with grant numbers C14/21/083 (J.V.) and C14/24/110 (K.T.). J.V. acknowledges the COST Action SUPERQUMAP (CA 21144) funding.

\pagebreak
\medskip

\renewcommand{\thefigure}{S\arabic{figure}}

\pagestyle{fancy}
\rhead{\includegraphics[width=2.5cm]{vch-logo.png}}

\title{Voltage-Controlled Dielectric Cavity Magnonics in Epitaxial Fe Films on SrTiO$_3$ Single Crystal Substrates : Supplementary Information}

\maketitle

\author{Stijn Reniers$^{1}$*}
\author{Ekaterina Gorokh$^{1}$}
\author{Lukas Nulens$^{1}$}
\author{Andries Boelen$^{3,4}$}
\author{Christian Haffner$^{4}$}
\author{Clement Merckling$^{4,3}$}
\author{Alejandro V. Silhanek$^{2}$}
\author{Kristiaan Temst$^{1,4}$}
\author{Joris Van de Vondel$^{1}$*}

\begin{affiliations}

$^{1}$ Quantum Solid-State Physics, Department of Physics and Astronomy, KU Leuven, Celestijnenlaan 200D, Leuven B-3001, Belgium

$^{2}$Department of Materials Engineering (MTM), KU Leuven, B-3001 Leuven, Belgium

$^{3}$Imec, Kapeldreef 75, Leuven, Belgium

$^{4}$Experimental Physics of Nanostructured Materials, Q-MAT, Department of Physics, Université de Liège, Liège B-4000, Belgium

*stijn.reniers@kuleuven.be
*joris.vandevondel@kuleuven.be
\end{affiliations}

\title{Voltage-Controlled Dielectric Cavity Magnonics in Epitaxial Fe Films on SrTiO$_3$ Single Crystal Substrates: Supplementary Information}

\section{Deposition and structural characterization} \label{StructSupp}

\begin{figure*}[h!]
    \centering
    \includegraphics[width=\linewidth]{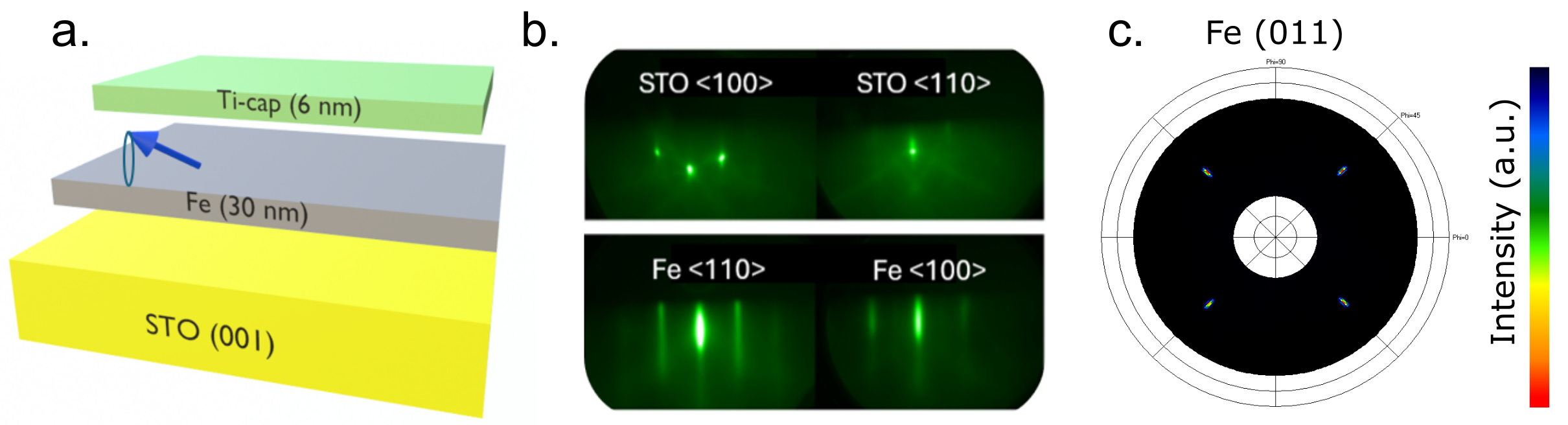}
    \caption{(a) STO/Fe sample stack with Ti-capping (b) RHEED-recordings of the bulk STO and Fe thin film surface for the STO/Fe magnonic cavity sample (c) Pole figure measurement of the epitaxial Fe film, measured on the (011)-reflection.}
    \label{fig:SuppStruct}
\end{figure*}

The epitaxial Fe film on STO substrate was structurally characterized during and after deposition. Figure \ref{fig:SuppStruct}a shows a schematic depiction of the sample stack, while Figure \ref{fig:SuppStruct}b,c show the electron- and X-ray diffraction characterization, respectively. Reflection High-Energy Electron Diffraction (RHEED, Figure \ref{fig:SuppStruct}b) shows clear diffraction spots and Kikuchi lines for the flat, crystalline STO surface, and pronounced streaks, indicating a crystalline and structurally continuous Fe surface during growth.
X-ray Diffraction (XRD) was performed as a pole figure measurement: the $\chi$-angle tilt (radial direction) was increased from 20 to 90 $^\circ$, while the the sample was rotated around its normal along the $\phi$-angle. The clear spots correspond to the (011) plane reflections when $\chi$ = 45$^\circ$.

\section{Volumetric magnetometry characterization  of the Fe film}\label{SQUIDSupp}

\begin{figure*}[h!]
    \centering
    \includegraphics[width=0.9\linewidth]{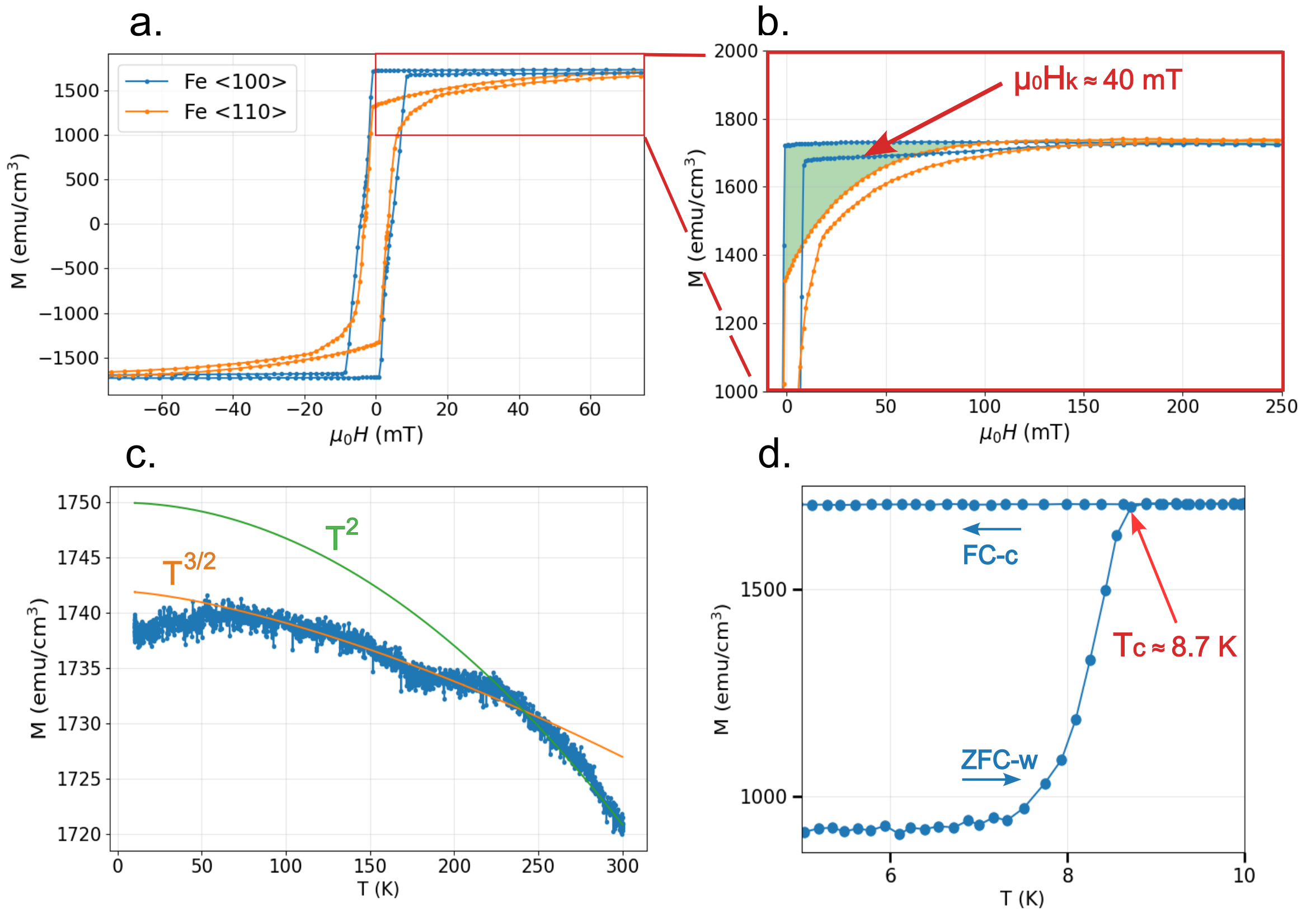}
    \caption{Volumetric magnetometry characterization by SQUID-VSM for in-plane magnetic field. (a) $M(H)$-measurement at $T=$ 10 K, along Fe $<$100$>$ (easy) and Fe $<$110$>$ (hard) directions. (b) Zoomed-in region on the first quadrant of the $M(H)$-measurement for the STO/Fe sample. Green shaded area corresponds to the magnetization work difference between easy and hard axes. This provides an estimate of the cubic anisotropy energy density $K_c$. (c) $M(T)$ measurement at $\mu_0H$ = 300 mT, recorded during cooling for the STO/Fe sample. (d) $M(T)$ measurement for the STO/Fe/Nb sample at $\mu_0H$ = 10 mT, yielding $T_c \approx$ 8.7 K.} 
    \label{fig:SuppSQUID}
\end{figure*}

To characterize the ferromagnetic properties of the epitaxial Fe film, volumetric magnetometry measurements were performed by SQUID-VSM. The sample volume was estimated as $( 5.0 \pm 0.5) \cdot 10^{-1} \times (3.0 \pm 0.3) \cdot 10^{-1} \times (30 \pm 2) \cdot 10^{-7}$ = $(45\pm 6)\cdot 10^{-8}$ cm$^3$. Supplementary Figure \ref{fig:SuppSQUID}a shows the M($H$)-loops along the Fe magnetocrystalline \textit{easy} $<$100$>$ and \textit{hard} $<$110$>$ axes, measured at $T=$10 K. The $M_s$-value based on the measured magnetic moment, normalized by the sample volume is (17 $\pm$ 2)$\cdot 10^2$ emu/cm$^3$ (kA/m). 

Figure \ref{fig:SuppSQUID}b shows a zoomed-in portion of the $M(H)$ curve in the first quadrant of the $M$-$H$ plane. The green shaded area corresponds to the difference in magnetizing work between the magnetocrystalline anisotropy (MCA) easy- and hard axes, equivalent to a quarter of the cubic magnetocrystalline anisotropy energy density $K_c$ \cite{CullityGraham}:
\begin{equation}
    W_{<100>}-W_{<110>} = \int_{0}^{M_s}(\mu_0H_{<100>}-\mu_0H_{<110>})dM = \frac{K_c}{4}
\end{equation}

This yields an estimate of $K_c$ = 32.2 kJ/m$^3$, corresponding to an anisotropy field $\mu_0H_k$ = 37 mT. For the simulations of the magnon-photon spectrum, we use a $\mu_0H_k$ value of 40 mT. Figure \ref{fig:SuppSQUID}c shows the $M(T)$ evolution, measured along the magnetocrystalline easy axis upon cooling in a quasi-saturation field of $\mu_0H$  = 300 mT.  The temperature evolution of magnetization is consistent with the evolution of the resonance line slope of the easy axis FMR measurement, shown as the blue curve in Figure 2e. The magnetization follows a $M_s(0)(1-BT^{2})$-law (green line) above $T$ = 200 K, and aligns better with the Bloch magnon $M_s(0)(1-BT^{3/2})$-relation \cite{Bloch1930Ferromagnetismus} (orange line) for lower $T$. This crossover in $M(T)$-dependence is consistent with experiments reported on Fe thin films \cite{Köbler_2002}.

\vspace{10 pt}

Supplementary Figure \ref{fig:SuppSQUID}d shows an $M(T)$ curve for the STO/Fe/Nb sample, recorded at $\mu_0H=$ 10 mT on warming (zero field-cooled) and subsequent cooling (field-cooled), along the Fe $<$100$>$-axis. The extracted superconducting critical temperature is $T_c \approx$  8.7 K. 

\newpage
\section{Dielectric permittivity evolution fit}\label{sup:perm}

\begin{figure*}[h!]
    \centering
    \includegraphics[width=0.7\linewidth]{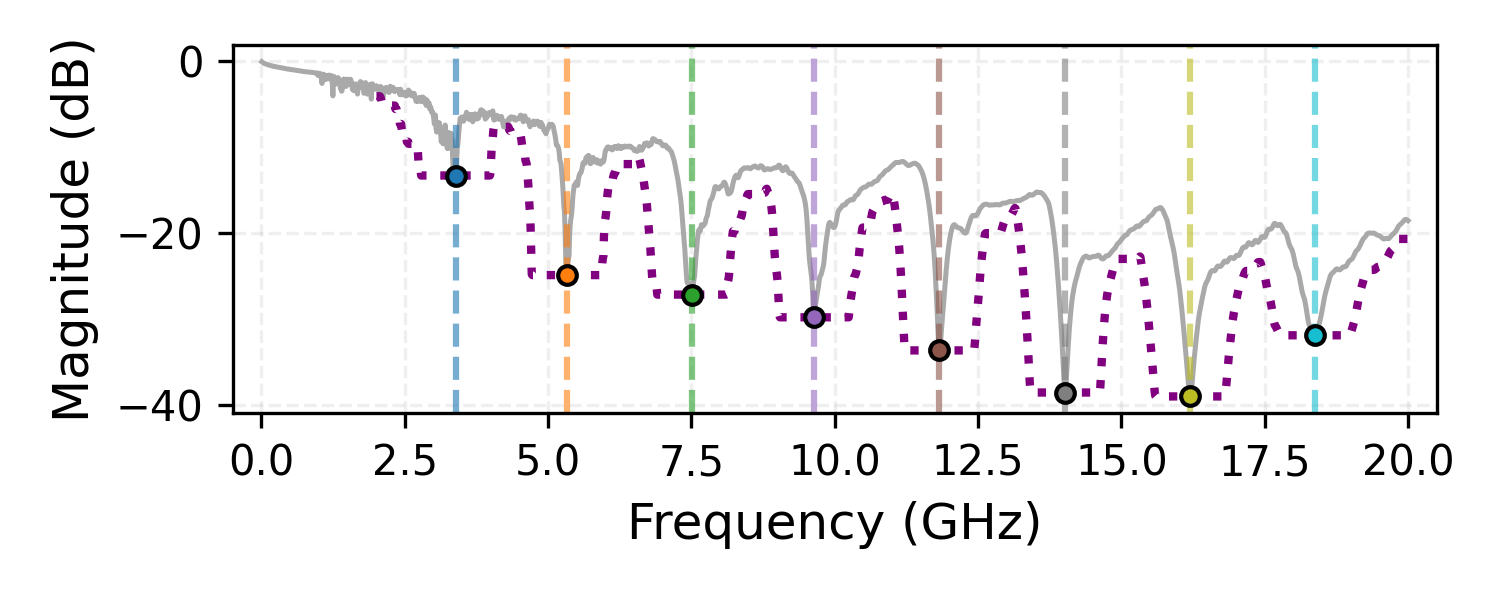}
    \caption{Transmission spectrum of the sample in the bare STO configuration at $T$ = 4 K and a driving power of 0 dBm (shown in grey). The eight observed out-of-plane $p$-modes are highlighted with colored markers and vertical dashed lines. The purple dotted line indicates the rolling-minimum floor envelope used for automated dip detection.}
    \label{fig:peak_detection_4K}
\end{figure*}

From the temperature-dependent microwave absorption spectrum of the bare STO sample measurement configuration in Figure 2a, the relative dielectric permittivity $\epsilon_r(T)$ can be quantitatively extracted. Figure \ref{fig:peak_detection_4K} shows a representative resonance spectrum measured at $T = 4$~K, where an automated peak-finding algorithm identifies transmission dips corresponding to distinct out-of-plane cavity modes ($p = 1$ to $p = 8$). Using the Fabry-P\'erot formula of Equation 1 in Subsection 3.1 of the main text, the effective microwave dielectric permittivity of STO as function of temperature $\epsilon_r(T)$ can be extracted from the temperature-dependent mode spectrum. 

\vspace{5 pt}
We use the phenomenological description given by Vendik and Zubko \cite{Vendik1997} to model the STO permittivity-dependence on temperature and electric field. In this model, $\epsilon_r(T,E)$ is described by the set of equations:

\begin{equation}
    \epsilon_r(T,E) = \epsilon_{00} \Phi(\xi, \eta)^{-1}
\end{equation}

\begin{equation}
    \Phi(\xi, \eta) = [\sqrt{\xi^2+\eta^3} + \xi]^{2/3} + [\sqrt{\xi^2+\eta^3} - \xi]^{2/3} - \eta
\end{equation}

\begin{equation}
    \xi = \sqrt{(E/E_N)^2 + \xi_{st}^2}
\end{equation}

\begin{equation}
    \eta = (\theta_F/T_c) \sqrt{1/16 - 0.01(\theta_F/T_c)+ (\theta_F/T_c)^2} -1
\end{equation}

Initial parameter values for $T_c$, $\epsilon_{00}$, $E_N$, $\theta_F$, $\xi_{st}$ were taken from \cite{Vendik1997}. 
Focusing on the dominant $p$-modes, the frequency separation $\Delta f = f_{p+1}-f_p$ was computed for all the fitted resonance peaks. The average, effective permittivity  $\epsilon_r(T)$ can then be found by the relation: 
\begin{equation}
    \epsilon_r = (\frac{c}{2L_z\Delta f})^2.
\end{equation} The resulting $\epsilon_r(T)$-dependence curve is shown in Supplementary Figure \ref{fig:Vendik_permittivity}a and Figure 2d from the main text. Optimal fit parameters for the Vendik-Zubko model were found to be $T_c$ = 39.6 K, $\epsilon_{00}$ = 3043, $\theta_F$ =  179.8 K and $\xi_{st}$ = 0.033. Using an internal normalization field $E_N$ = 11 kV/cm \cite{Vendik1997, Fujishita2016}, we additionally predict the impact of electric field on the $\epsilon_r(T)$-evolution. These results are shown in Figure \ref{fig:Vendik_permittivity}b.

\begin{figure*}[h!]
    \centering
    \includegraphics[width=\linewidth]{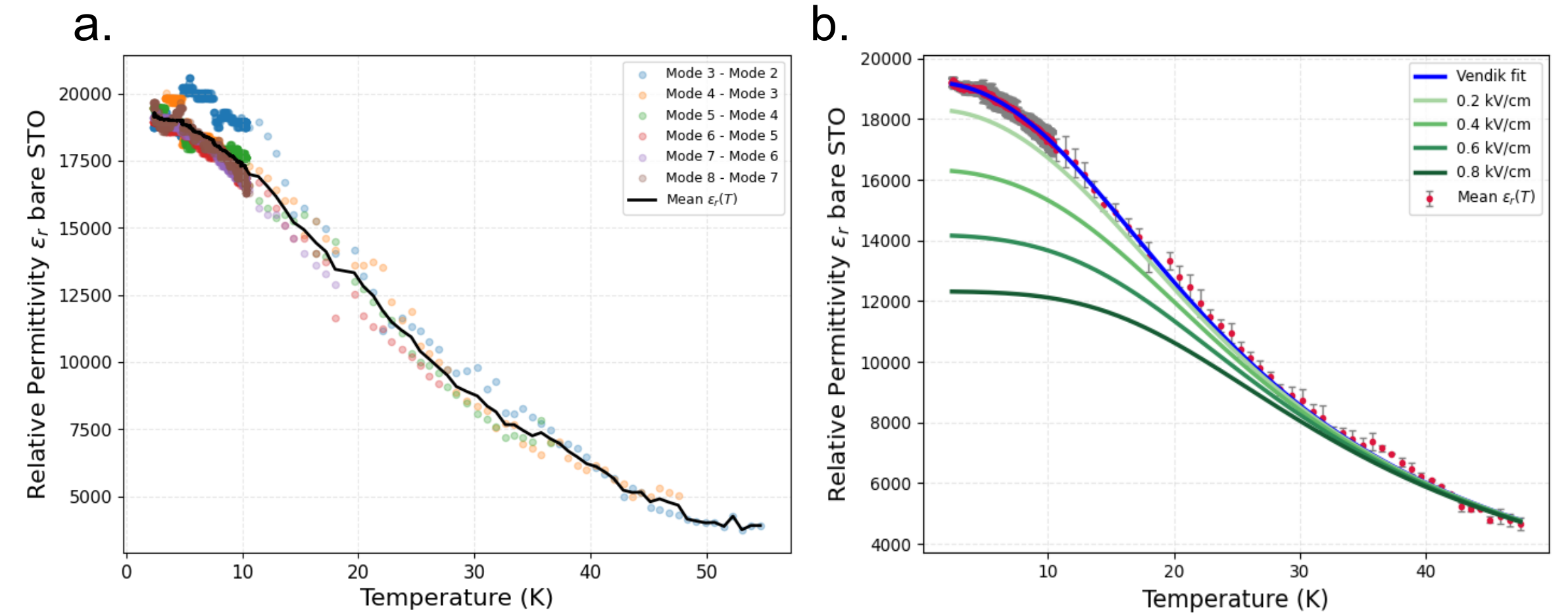}
    \caption{ (a) Average effective dielectric permittivity evolution $\epsilon_r(T)$, extracted from the resonance frequency spacings $\Delta f$ between the different pairs of $p$-modes below $T$ = 50 K. (b) Corresponding Vendik-Zubko model fit (blue) with the extracted model parameters, and a prediction for the impact of back-gate electric field (green curves).}
    \label{fig:Vendik_permittivity}
\end{figure*}

\newpage

\section{$S_{21}$ transmission parameter measurement at different $T$}\label{sup:heatmaps}

A more elaborated set of FMR heatmaps for different $T$-values was measured, aside from the ones shown in the main text. These are summarized in the following subsections.

\vspace{5 pt}
\subsection{Fe-side excitation}
\begin{figure*}[h!]
    \centering
    \includegraphics[width=0.9\linewidth]{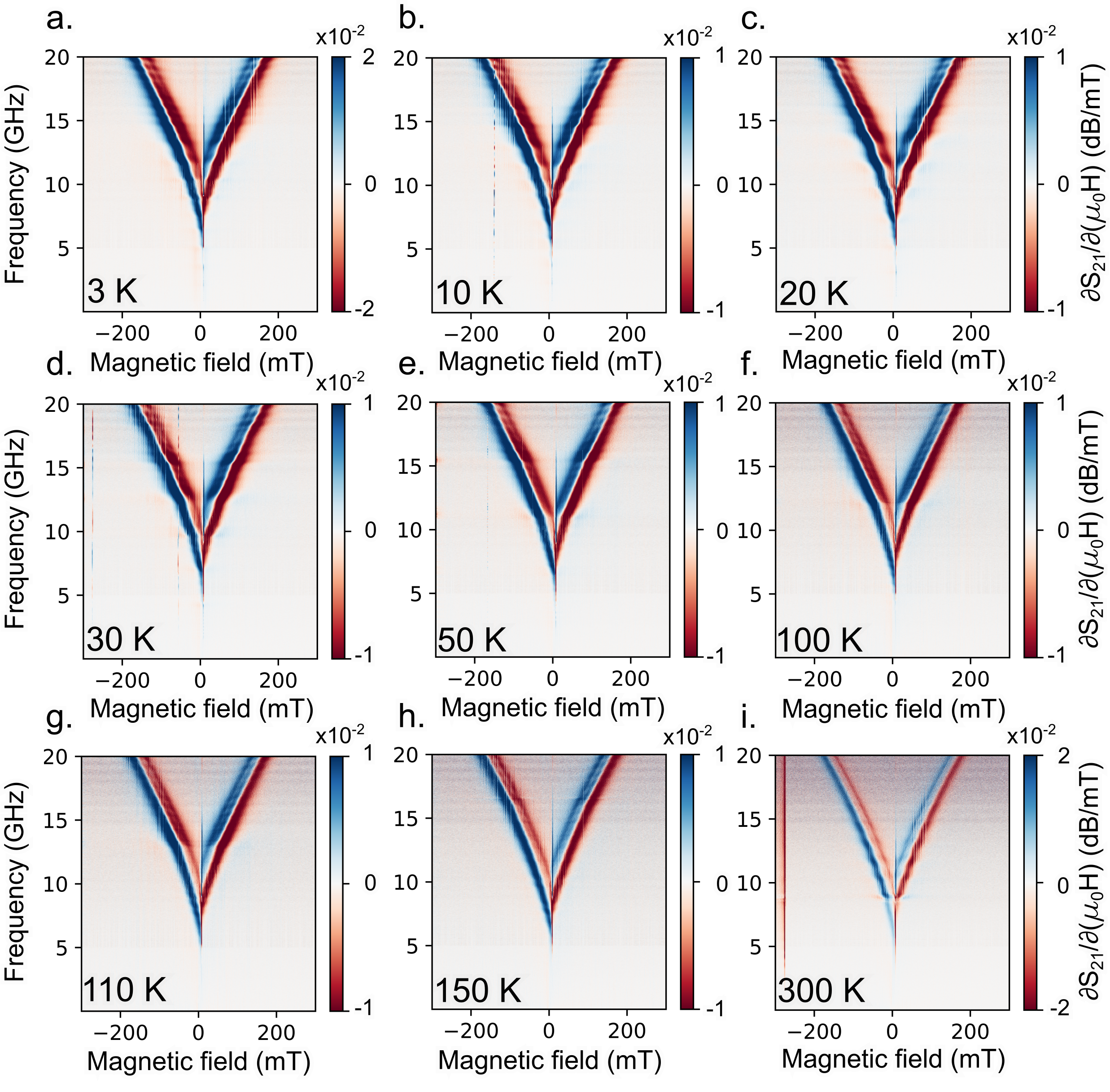}
    \caption{Easy Axis FMR spectra, recorded at different temperatures for the STO/Fe sample in Fe-side excitation configuration.}
    \label{fig:SuppFMR_EA}
\end{figure*}

Supplementary Figure \ref{fig:SuppFMR_EA} shows the field-dependent microwave response for Fe-side excitation configuration along the Fe$<$100$>$ magnetocrystalline easy axis. The local slope suppressions, accompanied by enhanced absorption intensity, demonstrate dissipative magnon-photon coupling, well-correlated with the temperature-dependence of the cavity mode frequencies, demonstrated in Figure 2b and Subsection 4.1 of the main text.

\vspace{10 pt}

To demonstrate the presence of magnetocrystalline anisotropy in the magnetization dynamics response, the Fe-side excitation was also measured along the Fe $<$110$>$ hard axis, shown in Supplementary Figure \ref{fig:SuppFMR_HA}. The initial negative slope of the resonance line is characteristic for the hard axis measurement, and the magnetic field of minimum resonance frequency lies close to the $\mu_0H_k \approx $ 40 mT anisotropy field estimate obtained by SQUID-VSM magnetometry, discussed in Supplementary Section \ref{SQUIDSupp}.

\begin{figure*}[h!]
    \centering
    \includegraphics[width=0.9\linewidth]{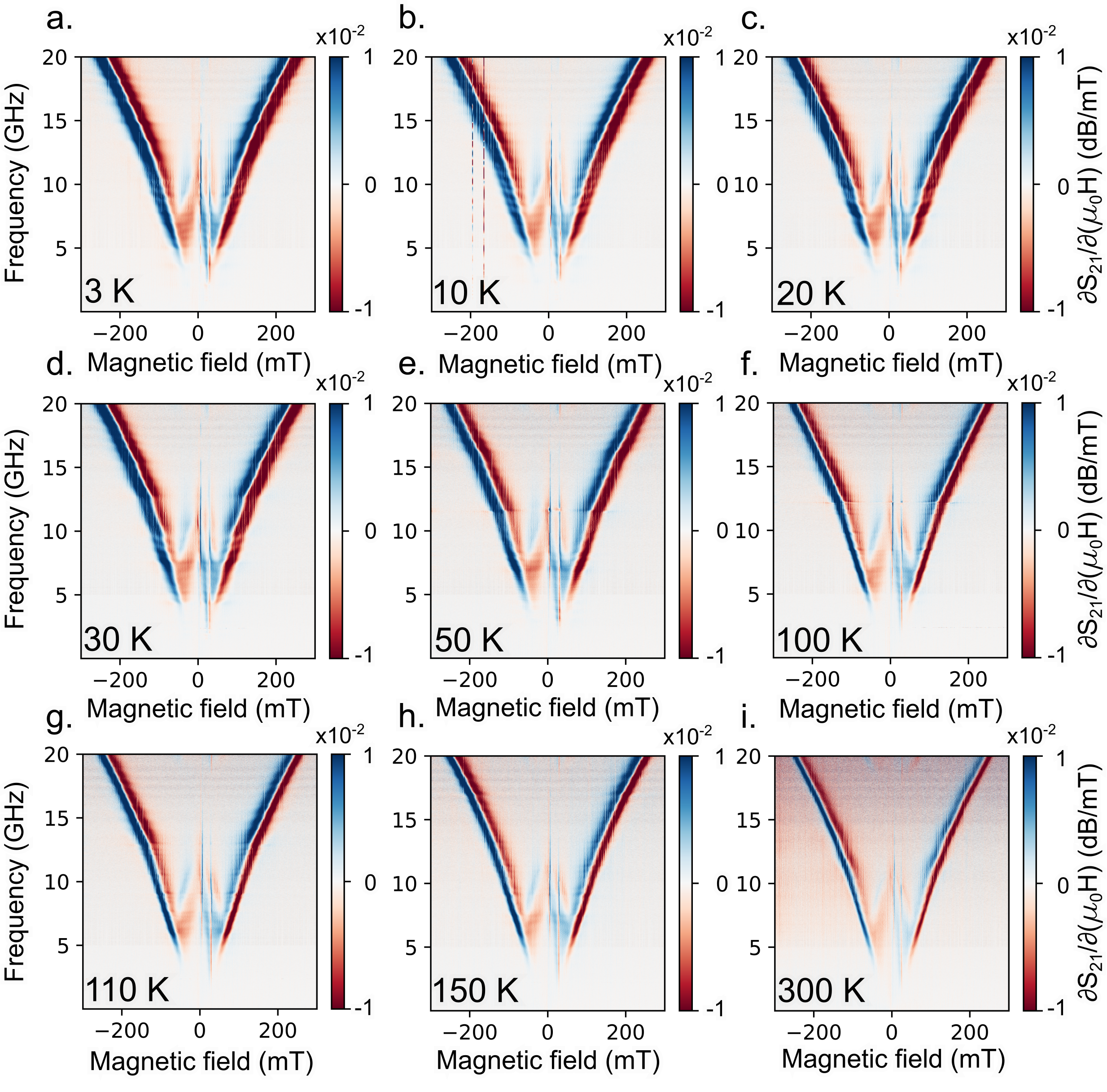}
    \caption{Hard Axis FMR spectra, recorded at different temperatures for the STO/Fe sample in Fe-side excitation configuration.}
    \label{fig:SuppFMR_HA}
\end{figure*}

\clearpage

\subsection{STO-side excitation}
\begin{figure*}[h!]
    \centering
    \includegraphics[width=0.9\linewidth]{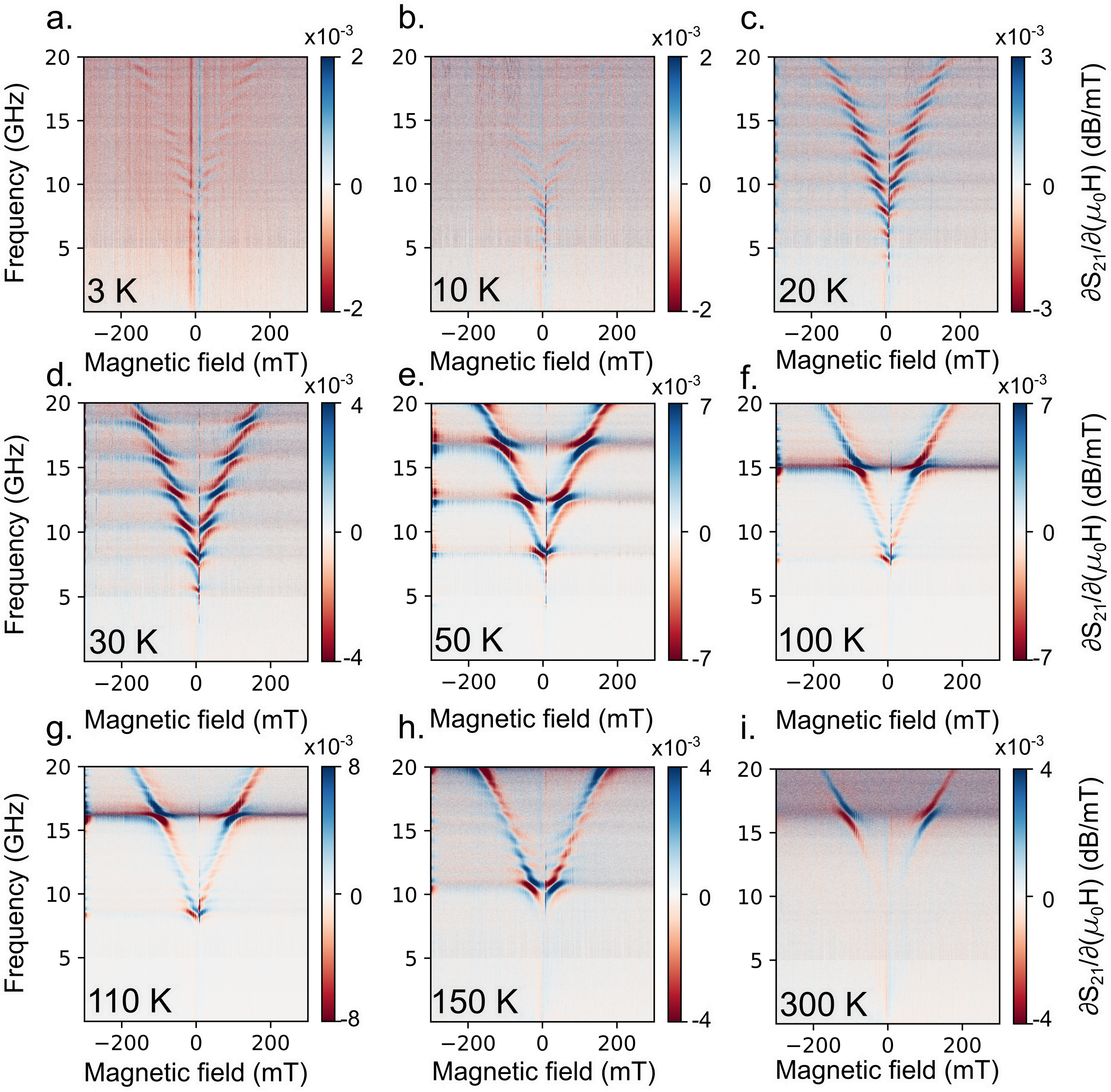}
    \caption{Easy Axis FMR spectra, recorded at different temperatures for the STO/Fe sample in STO-side excitation configuration.}
    \label{fig:SuppFMR_EA_bottom}
\end{figure*}

Supplementary Figure \ref{fig:SuppFMR_EA_bottom} shows the field-dependent microwave response for STO-side excitation configuration, in the coherent coupling regime. The local avoided crossings demonstrate predominantly coherent magnon-photon coupling, well-correlated with the cavity mode frequencies temperature-dependence, demonstrated in Figure 2c and Subsection 4.2 of the main text.

As shown in the low-temperature FMR spectra in Supplementary Figure \ref{fig:SuppFMR_EA_bottom}, the FMR response is strongly suppressed at temperatures below $T=$ 20 K. To improve the intensity of the mode spectrum, a thicker metallic layer is required. We therefore selected superconducting Nb, with a critical temperature of $T_c \approx 8.7$ K (see Supplementary Figure \ref{fig:SuppSQUID}d). The hybrid mode spectrum is sharper and more intense with the Nb layer, both above and below the superconducting transition temperature of Nb, shown in Supplementary Figure \ref{fig:SuppFMR_Nb}). As briefly discussed in Subsection 4.2 of the main text, and reported in \cite{GolovchanskiySFHybrid, AFMGolovchanskiy, golovchanskiy2019interplay}, the mode spectrum at $T=$ 5 K appears to show an additional, higher order magnon branch. While not investigated within this work, superconducting materials can provide additional tuning of the magnon modes.

\begin{figure*}[h!]
    \centering
    \includegraphics[width=\linewidth]{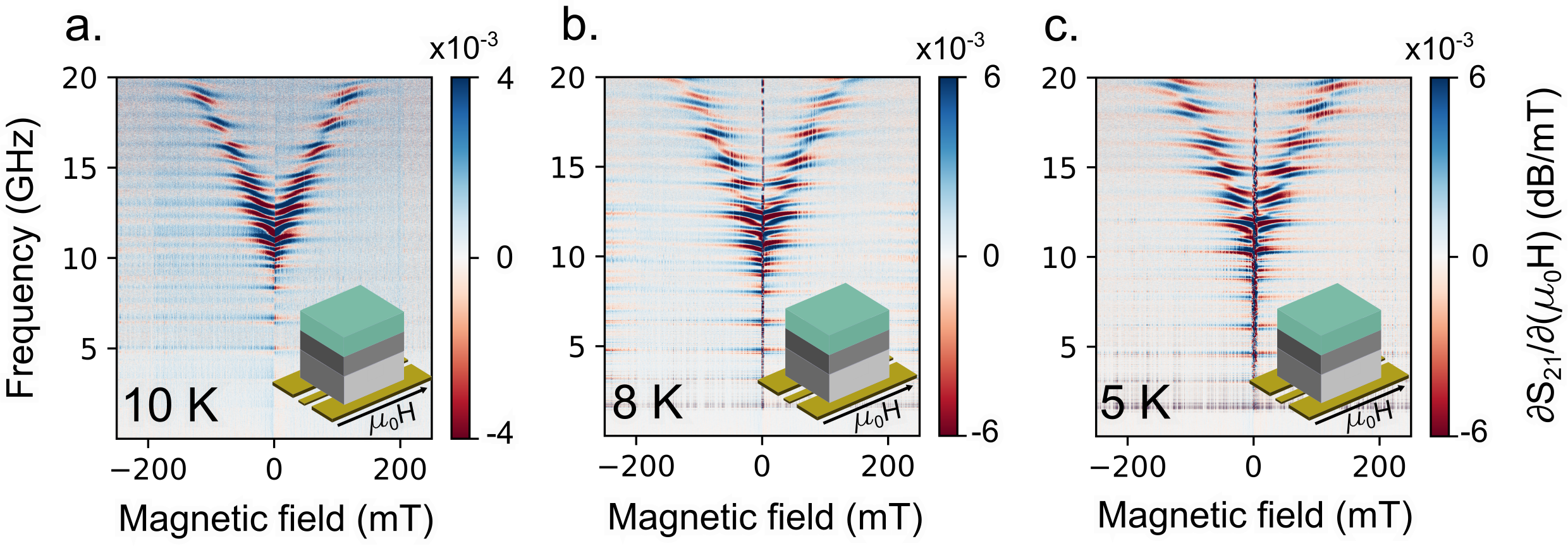}
    \caption{(a-c) FMR signal of STO/Fe/Nb  system at $T=$ 10 K, 8 K and 5 K, respectively.}
    \label{fig:SuppFMR_Nb}
\end{figure*}

\subsection{Phenomenological coupling Hamiltonian and $S_{21}$ simulations}\label{sup:coupcoef}

The quantitative modeling of the hybrid mode spectrum, outlined in Subsection 4.3 of the main text, was implemented in Python. The Vendik-Zubko equations were used, with the parameters fitted in Supplementary Section \ref{sup:perm}, to compute the STO permittivity as function of temperature, while the $k=0$ magnon mode was computed using the Kittel equation, with an anisotropy field estimate $\mu_0H_k$ = 40 mT. 

\vspace{10 pt}

\begin{figure*}[h!]
    \centering
    \includegraphics[width=0.7\linewidth]{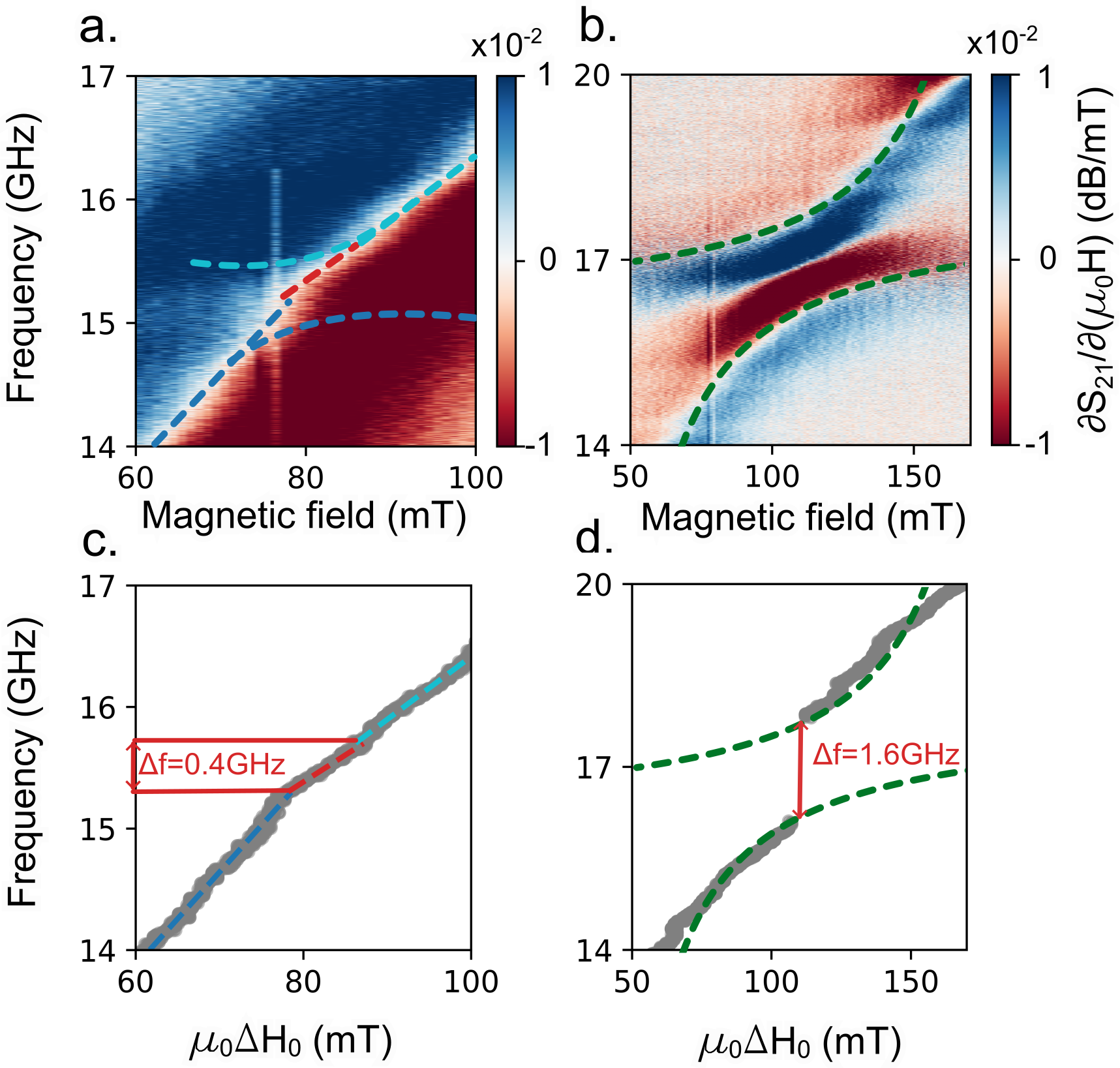}
    \caption{Extracted coupling strengths at $T$ = 50 K for (a) \textit{Fe-side excitation}, where dissipative coupling is characterized by \textit{level attraction} and an associated slope change. (b) \textit{STO-side excitation}, where coherent coupling is characterized by \textit{avoided crossing} and an associated opening of the hybrid mode spectrum. (c) and (d) show the extracted $\Delta$f values for dissipative and coherent coupling, respectively. The dashed lines provide a guide to the eye for the mode splitting.}
    \label{fig:SuppFMRcoupling}
\end{figure*}

The coupling Hamiltonian matrix was then constructed using all cavity and magnon modes, and diagonal elements were computed based on the coupling strength estimates obtained in Subsection 4.3 of the main text. Supplementary Figure \ref{fig:SuppFMRcoupling} shows a more detailed view of the parameter extraction procedure (shown in Figure 5 of the main text) at $T = 50$ K.

Using these representative estimates for the magnon-photon coupling strength, the hybridized mode spectrum in \textit{dissipative} coupling configuration (Fe-side excitation) was first modeled for the three temperatures discussed in the main text. The result is shown in Supplementary Figure \ref{fig:SuppFMR_modelVendik}. 

\begin{figure*}[h!]
    \centering
    \includegraphics[width=\linewidth]{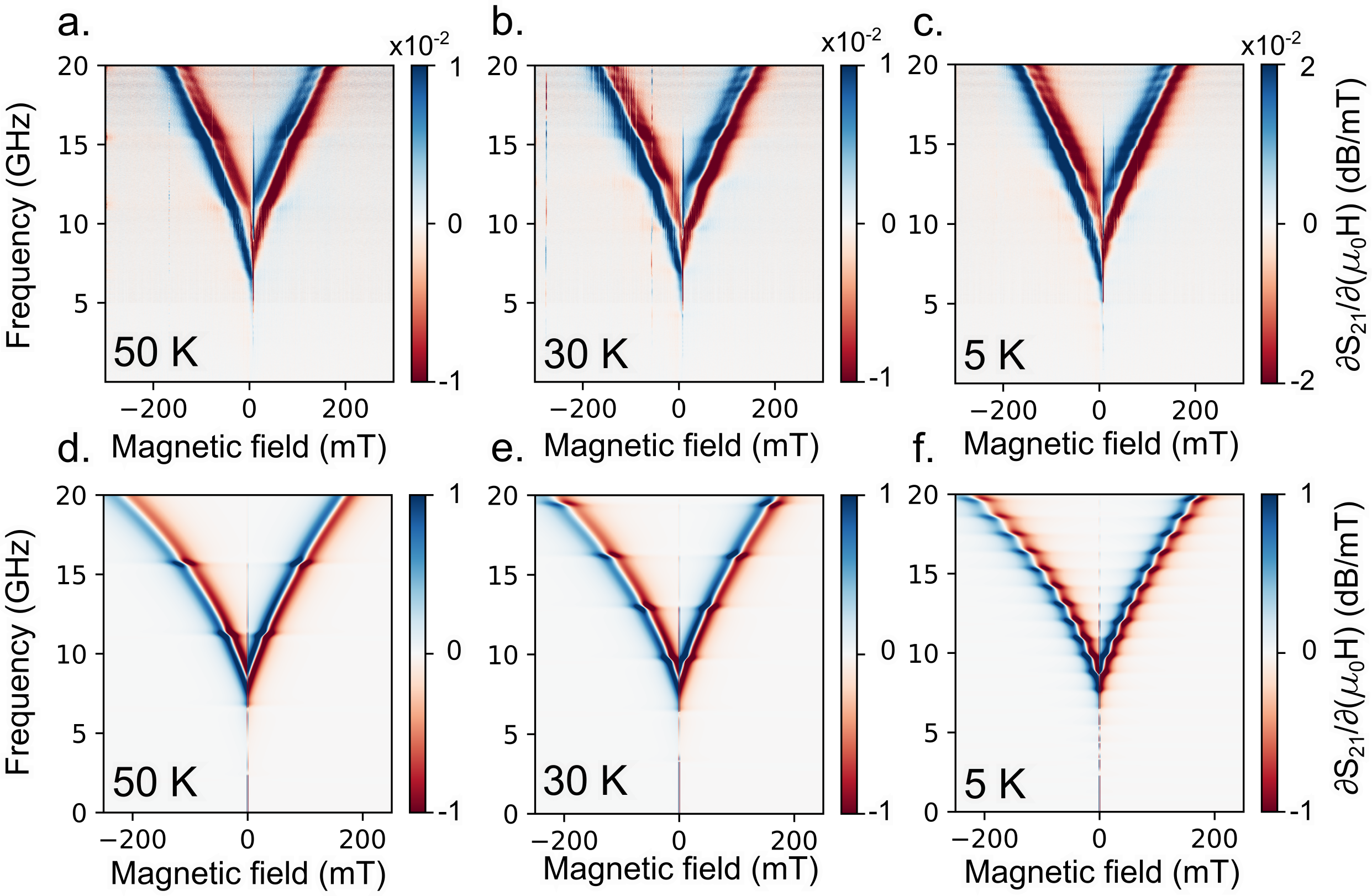}
    \caption{Phenomenological simulation results based on the coupled-mode Hamiltonian and Vendik-Zubko model of STO permittivity, compared to the experimental results. Top row (a-c) shows the experimental data for Fe-side excitation configuration at $T$ = 50 K, 30 K and 5 K. Bottom row (d-f) shows the simulated spectra for Fe-side excitation configuration at $T$ = 50 K, 30 K and 5 K. }
    \label{fig:SuppFMR_modelVendik}
\end{figure*}

The hybridized mode spectrum in \textit{coherent} coupling configuration (STO-side excitation) was then also modeled for the three temperatures discussed in the main text. The result is shown in Supplementary Figure \ref{fig:SuppFMR_modelVendik_coherent}.

\begin{figure*}[h!]
    \centering
    \includegraphics[width=\linewidth]{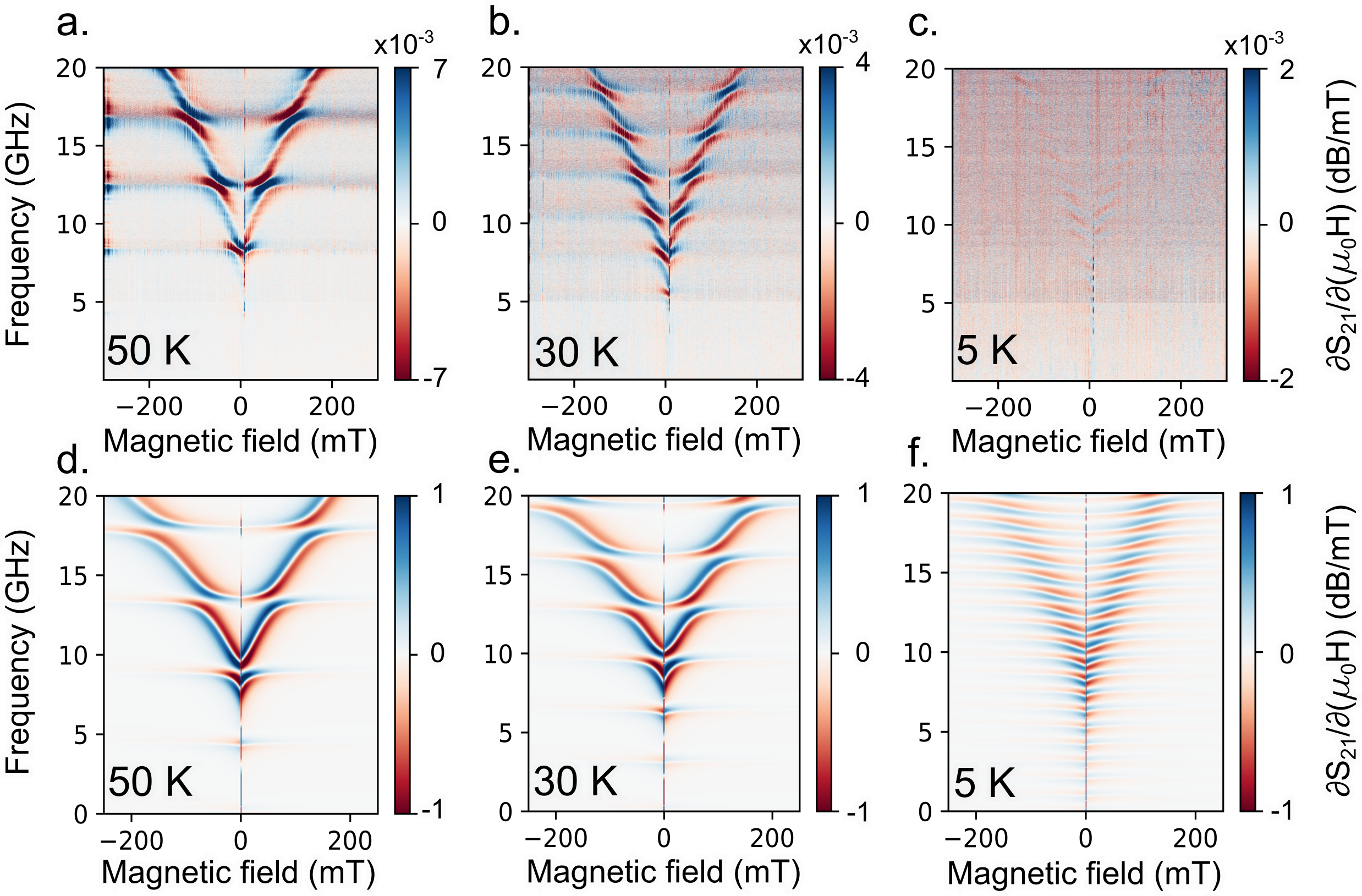}
    \caption{Phenomenological simulation results based on the coupled-mode Hamiltonian and Vendik-Zubko model of STO permittivity, compared to the experimental results. Top row (a-c) shows the experimental data for STO-side excitation configuration at $T$ = 50 K, 30 K and 5 K. Bottom row (d-f) shows the simulated spectra for STO-side excitation configuration at $T$ = 50 K, 30 K and 5 K. }
    \label{fig:SuppFMR_modelVendik_coherent}
\end{figure*}

\vspace{10 pt}
The changes in FMR intensity in the frequency region where the interaction with the STO modes occurs, as discussed in Section 4 of the main text, are shown in Supplementary Figure \ref{fig:SuppFMR_model_intensity}. The dip intensity increases below the STO resonance frequencies, whereas the peak intensity increases above them. This behavior is clearly visible in the intensity modulation extracted from the experimental data and is also reproduced by the Vendik–Zubko model. The frequency separation between the dip- and peak-intensity enhancements is smaller in the simulated data because the modeled FMR spectrum has higher quality factors and idealized, Lorentzian mode shapes.

\begin{figure*}[h!]
    \centering
    \includegraphics[width=\linewidth]{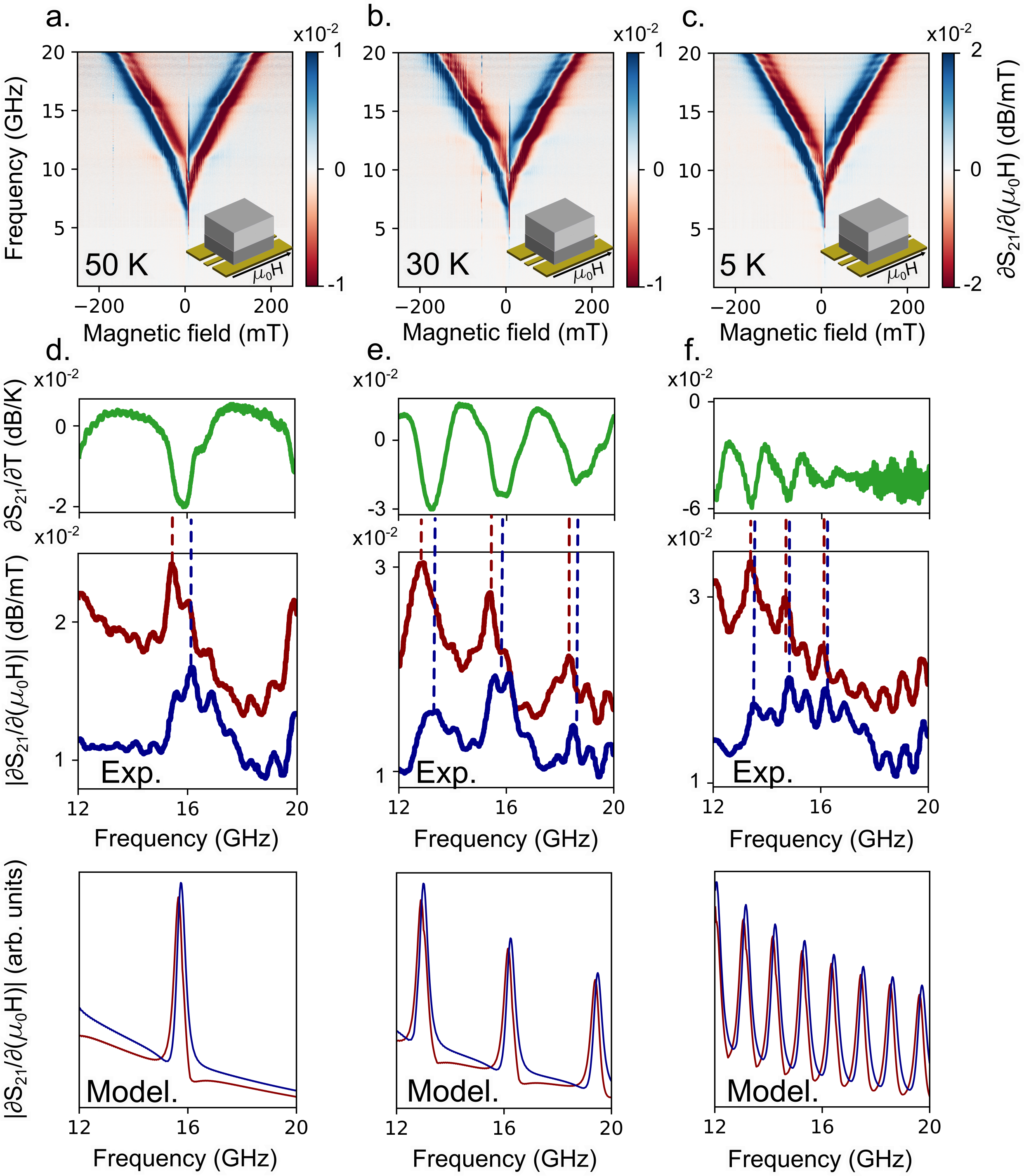}
    \caption{(a-c) Ferromagnetic resonance characterization of the STO (gray)/Fe (dark gray) system at different temperature values ($T$ = 50 K, 30 K and 5 K), in the form of $\partial S_{21}$($f,\mu_0H$)/$\partial (\mu_0H)$ heatmaps. The left axis on panels a-c corresponds to the RF frequency, while the right axis color bar indicates the intensity of the $S_{21}$ partial derivative with respect to magnetic field. (d-f) Correlation of the FMR peak/dip intensities (dark blue/red) (obtained from both the experimental data and model) with the cavity resonance spectrum $\partial S_{21}(f,T)/\partial T$ (green) from the main text Figure 2b.}
    \label{fig:SuppFMR_model_intensity}
\end{figure*}

\clearpage

\section{Voltage-controlled magnon-photon coupling measurements}\label{sup:gatedependent}

\begin{figure*}[h!]
    \centering
    \includegraphics[width=0.6\linewidth]{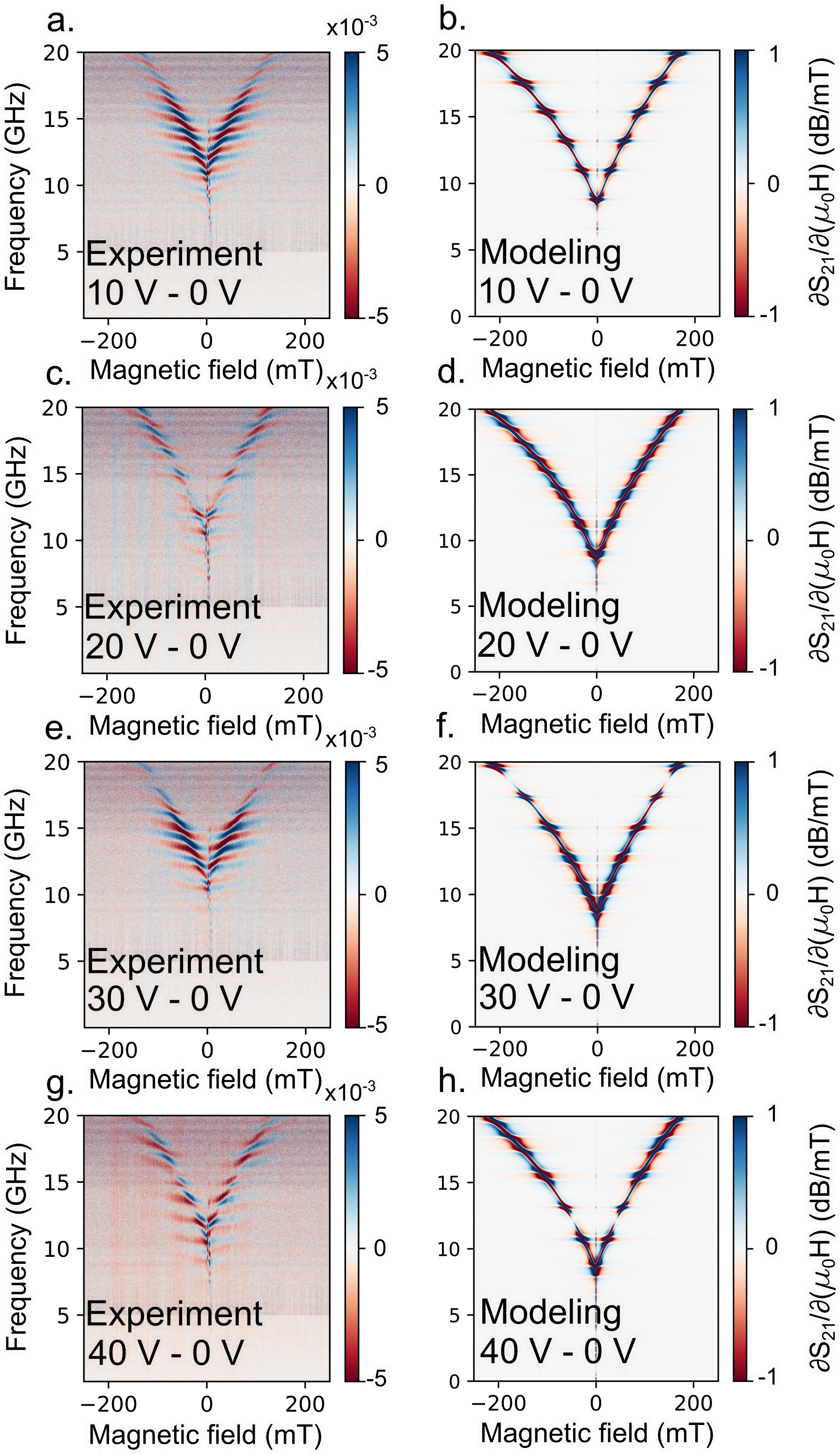}
    \caption{Experimental (left) and simulated (right) difference maps, with respect to $V_G$ = 0 V, of the gated  $\partial S_{21}$($f,\mu_0$H)/$\partial (\mu_0H$)-spectrum, for all positive applied gate voltages, assuming $f = n\frac{\lambda}{2}$-resonance modes only.} \label{fig:SuppFMR_modelVendik_gating}
\end{figure*}

Electric field control of the hybrid mode spectrum was modeled by incorporating the Vendik-Zubko model for electric field-dependence of the STO permittivity, and the related dielectric cavity modes. Supplementary Figures \ref{fig:SuppFMR_modelVendik_gating} a,c,e,g show the experimentally determined difference spectra as function of gate voltage (compared to 0 V), correlating well with the results of Section 5 of the main text (Figure 6). Supplementary Figures \ref{fig:SuppFMR_modelVendik_gating} b,d,f,h show the simulated difference maps for gate voltages 10 V, 20 V, 30 V and 40 V, with respect to the 0 V mode spectrum at $T$ = 4 K.

The dependence of the STO resonance modes on the electric field, plotted in Figure 6e of the main text, for a larger range of the frequencies ($f=$ 8-20 GHz) is shown in Supplementary Figure \ref{fig:SuppFMR_gating1}a. The FMR dip depth (absolute value) shown in Figure 6f of the main text, for a larger range of frequencies ($f=$ 8-20 GHz) is shown in Supplementary Figure \ref{fig:SuppFMR_gating1}b. The FMR slope modulation due to the interaction with the STO resonance mode for a smaller range of frequencies (12-16 GHz) is shown in Supplementary Figure \ref{fig:SuppFMR_gating1}c.


\begin{figure*}[h!]
    \centering
    \includegraphics[width=\linewidth]{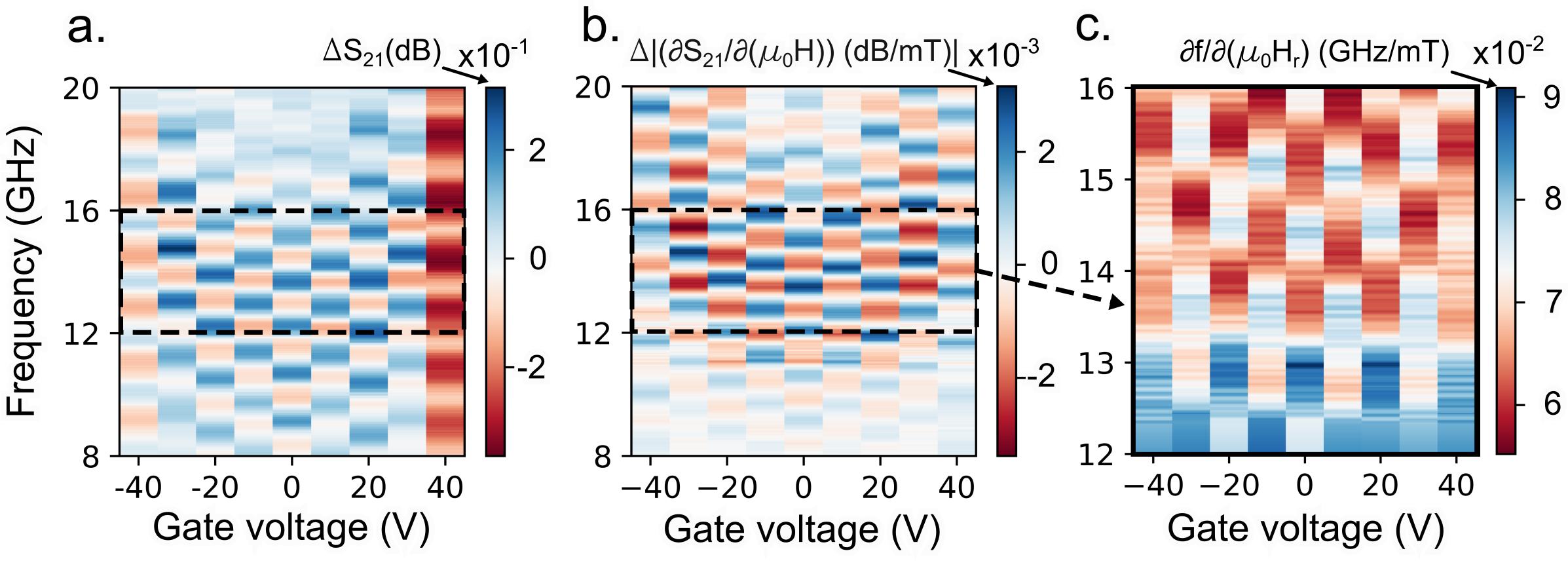}
    \caption{(a) Raw FMR data measured at $\mu_0H$ = 250 mT, far outside the FMR resonance condition for 8-20 GHz. The average transmission over all gate voltages was subtracted as baseline, emphasizing the voltage-induced shifts of the STO resonances. (b) Dip depth in the FMR outer branch as a function of gate voltage for different frequencies for 8-20 GHz. The heatmap shows the data after baseline subtraction of the average over all gate voltages, to remove the background. (c) Change in the FMR slope within the narrower frequency range, indicated by the black dashed box in panels a,b.}
    
    \label{fig:SuppFMR_gating1}
\end{figure*}

The STO resonance modes, used for the plotting of the heatmap, shown in Supplementary Figure \ref{fig:SuppFMR_gating1}a, are shown in Supplementary Figure \ref{fig:SuppFMR_gating2}a,b for both negative and positive gate voltages, respectively.

\begin{figure*}[h!]
    \centering
    \includegraphics[width=0.7\linewidth]{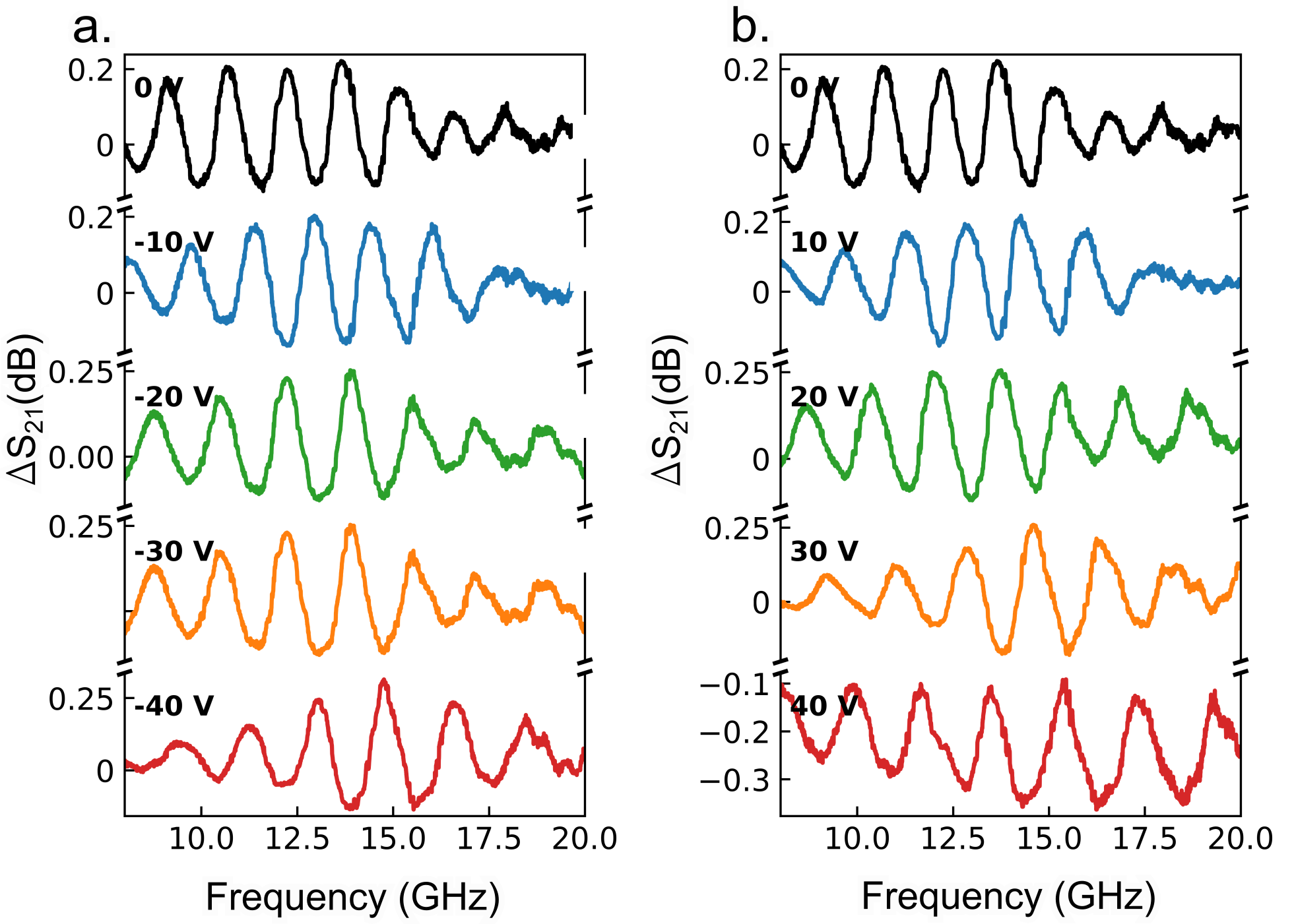}
    \caption{(a) STO cavity mode spectrum cross-sections for negative gate voltages. (b) STO cavity mode spectrum cross-sections for positive gate voltages.}
    \label{fig:SuppFMR_gating2}
\end{figure*}

\medskip
\newpage\clearpage
\bibliographystyle{MSP}
\bibliography{Ref.bib}

\end{document}